\documentclass[a4paper,12pt]{article}

\usepackage{graphicx, amsmath, amsfonts, amssymb, array, natbib, graphicx, float, lipsum, tikz, framed}

\tikzset{
  treenode/.style = {shape=rectangle,
                     draw, align=center,
                     top color=white, text height = 0.5cm},
  root/.style     = {treenode, font=\Large},
  env/.style      = {treenode, font=\Large},
  dummy/.style    = {shape=circle,draw,align=center, top color=white, font=\large}
}

\title{Predicting human-driving behavior to help driverless vehicles drive: random intercept Bayesian Additive Regression Trees}
\author{Yaoyuan V. Tan, Carol A.C. Flannagan, Michael R. Elliott}

\begin{document}
\maketitle
\linespread{2}
\selectfont

\begin{abstract}
The development of driverless vehicles has spurred the need to predict human driving behavior to facilitate interaction between driverless and human-driven vehicles. Predicting human driving movements can be challenging, and poor prediction models can lead to accidents between the driverless and human-driven vehicles. We used the vehicle speed obtained from a naturalistic driving dataset to predict whether a human-driven vehicle would stop before executing a left turn. In a preliminary analysis, we found that BART produced less variable and higher AUC values compared to a variety of other state-of-the-art binary predictor methods. However, BART assumes independent observations, but our dataset consists of multiple observations clustered by driver. Although methods extending BART to clustered or longitudinal data are available, they lack readily available software and can only be applied to clustered continuous outcomes. We extend BART to handle correlated binary observations by adding a random intercept and used a simulation study to determine bias, root mean squared error, 95\% coverage, and average length of 95\% credible interval in a correlated data setting. We then successfully implemented our random intercept BART model to our clustered dataset and found substantial improvements in prediction performance compared to BART and random intercept linear logistic regression.  

\end{abstract}


\section{Introduction}

In transportation statistics, a new area of research brought about by improvements in artificial intelligence and engineering is the creation of the autonomous (self-driving) vehicle. These vehicles have been tested on city streets in certain locations since 2009. A number of companies have deployed or announced plans for deployment of such vehicles \citep{google,tesla,gm}. A major hurdle for self-driving vehicles on public roads is that these vehicles will have to interact with human-driven vehicles for the foreseeable future. Human drivers do not always communicate their plans to other drivers well. For example, when making a turn, the turn signal is the only explicit means of communicating plans, and even they are used with less than perfect reliability. Hence, the ability to deploy driverless vehicles on a large scale will critically depend on the development of a good prediction model for human driving behavior.   

Building a prediction model that addresses all or most of the human driving behavior is a massive and complex task. To keep this paper concise, we focus on the the development of a prediction model for a single driving behavior: whether a human driver would stop at an intersection before executing a left turn. We are particularly interested in left turn stops because in countries with right-side driving, for example, US, left turn crashes can result in severe passenger-side impacts. Since left turn maneuvers already present a challenge for human drivers, we expect this maneuver to present difficulty for the driverless vehicle. 

To develop our prediction model, we used a naturalistic driving study, the Integrated Vehicle Based Safety System (IVBSS) study \citep{sayer}. Naturalistic driving studies (including the IVBSS) involve the collection of driving data from vehicles as they are piloted on actual roads. These driving data are collected by a data acquisition system (DAS) installed on a study subject's vehicle or a research vehicle. Typical data collected include vehicle speed, brake application, and miles traveled.

Prediction models in statistics typically rely on regression models that require estimation of covariate main effects and interactions, and, when predictors are continuous or on a fine ordinal scale, assessment of non-linearities. In the settings where understanding associations or, under appropriate assumptions, causal mechanism between predictors and outcomes are of interest, approximations for non-linearities and averaging over interactions might be used to develop summaries to ease interpretation. In prediction, since obtaining the most accurate forecast is the goal, estimating highly complex non-linearities, including the interactions, is at a premium, as long as these non-linearities are true signals and not noise. 

Perhaps the most common method for modeling non-linearity is to use a polynomial transformation for a covariate, usually centered at the mean to reduce correlation. More sophisticated approaches use penalized splines or additive models that only require assumptions of smoothness (existence of derivatives) to obtain consistent estimates of a non-linear trend \citep{hastie_book, ruppert}. Modeling of non-linear interactions between two or more predictors using thin-plate splines \citep{franke} can quickly become difficult, suffering from the ``curse of dimensionality'', as the data required to estimate high-dimensional surfaces become enormous. In the binary outcomes setting, methods such as classification and regression trees \citep[CART;][]{breiman} as well as more sophisticated machine learning techniques such as artificial neural networks \citep[ANN;][]{smith} and support vector machines \citep[SVM;][]{gammermann} are commonly used. Although CART is able to model complex interactions naturally, it faces difficulty when modeling non-linear interactions. In contrast, ANN and SVM excel at modeling non-linearities but may face difficulties when modeling complex interactions.

Because our goal is prediction, we prefer regression methods that are able to account for non-linear main and multiple-way interaction effects. Based on preliminary analyses, we chose the Bayesian additive regression trees (BART) \citep{chipman_bart} to predict whether a human-driven vehicle would stop before executing a left turn at an intersection. Because BART was designed for independent subjects, we needed to extend BART to take into account the clustering in our dataset. We are aware of two papers that extended BART to handle longitudinal or clustered observations: \cite{zhang} used a spatial random intercept BART to merge two datasets, and \cite{lowkam} did so in a dose-finding toxicity study. \cite{zhang} developed an imputation model for a statistical matched problem \citep{rassler} that used BART with a conditional auto-regressive distribution for the random intercept. Since the correlation our dataset was induced by repeated measurements and not spatial effects, the distribution Zhang et al. placed on the random intercept may not be appropriate. Moreover, they did not discuss how their model could be extended to clustered binary outcomes. \cite{lowkam} investigated the associations between the physico-chemical properties of nanoparticles and their toxicity profiles over multiple doses. The complex nature of their goal prompted them to first specify an autoregressive covariance matrix with truncated support on $[0,1]$ to handle the correlated measurements, and then they specified a conditionally conjugate P-spline prior for the terminal nodes of the regression trees. The complexity of their method makes implementation to our dataset difficult since our outcomes are binary. Neither papers provided convenient software for implementing their methods.

Motivated by the lack of an appropriate and straightforward method to implement BART to handle clustered binary outcomes, we propose an extension of BART to account for longitudinal binary observations. Our proposed method accounts for clustering by adding a random intercept to BART and we call this random intercept BART (riBART). In the next section, we provide a review of BART. In Section 3, we present riBART followed by a simulation study in Section 4 to compare the performance of BART and riBART when applied to clustered datasets. In Section 5, we implement riBART on our dataset and compare its prediction performance with BART, linear logistic regression, and random intercept linear logistic regression. Finally, we conclude with a discussion and possible future work in Section 6.

\section{Bayesian Additive Regression Trees}
\subsection{Continuous outcomes}
Denote a continuous outcome $Y_k$ with associated $p$ covariates $\mathbf{X}_k=(X_{k1},\ldots,X_{kp})^T$ for $k=1,\ldots,n$ subjects. BART models the outcome as
\begin{equation}
	\label{bart_main_con}
	Y_k=\sum_{j=1}^m g(\mathbf{X}_k,T_j,\mathbf{M}_j)+\epsilon_k \quad\epsilon_k\overset{i.i.d.}{\sim} N(0,\sigma^2)
\end{equation}
where $T_j$ is the $j^{\text{th}}$ binary tree structure and $\mathbf{M}_j=(\mu_{1j},\ldots,\mu_{b_jj})^T$ is the set of $b_j$ terminal node parameters associated with tree structure $T_j$ \citep{chipman_bart}. $g(\mathbf{X}_k,T_j,\mathbf{M}_j)$ can be viewed as the $j^{\text{th}}$ function that assigns the mean $\mu_{ij}$ to the $k^{\text{th}}$ outcome, $Y_k$. Typically, the number of trees $m$ is fixed and no prior distribution is placed on $m$. Chipman et al. suggested setting $m=200$ as this performs well in many situations. Alternatively, cross-validation could be used to determine $m$ \citep{chipman_bart}.

The joint prior distribution for (\ref{bart_main_con}) is $P[(T_1,\mathbf{M}_1),\ldots,(T_m,\mathbf{M}_m),\sigma]$. Note that by the independence of $\epsilon_k$ and $(T_j,\mathbf{M}_j)$ as well as the independence between all $m$ tree structures and terminal node parameters, the joint prior distribution $P[(T_1,\mathbf{M}_1),\ldots,(T_m,\mathbf{M}_m),\sigma]$ can be decomposed as
\begin{align*}
	P[(T_1,\mathbf{M}_1),\ldots,(T_m,\mathbf{M}_m),\sigma]&=[\prod_{j=1}^mP(T_j,\mathbf{M}_j)]P(\sigma)\\
		&=[\prod_{j=1}^mP(\mathbf{M}_j|T_j)P(T_j)]P(\sigma)\\
		&=[\prod_{j=1}^m\{\prod_{i=1}^{b_j}P(\mu_{ij}|T_j)\}P(T_j)]P(\sigma).
\end{align*}
where $i=1,\ldots,b_j$ indexes the terminal node parameters in tree $j$. This implies that we need to assign priors to $T_j$, $\mu_{ij}|T_j$, and $\sigma$ in order to obtain the posterior distributions of $T_j$, $\mu_{ij}$, and $\sigma$. \cite{chipman_bart} suggested the following  prior distributions on $\mu_{ij}|T_j$ and $\sigma$:
\begin{align*}
	\mu_{ij}|T_j&\sim N(\mu_{\mu},\sigma^2_{\mu}),\\
	\sigma^2&\sim IG(\frac{\nu}{2},\frac{\nu\lambda}{2}).
\end{align*} 
where $IG(\alpha,\beta)$ is the inverse gamma distribution with shape parameter $\alpha$ and rate parameter $\beta$. The prior distribution of $P(T_j)$ can be specified using three aspects: (i) the probability that a node at depth $d=0,1,2,\ldots$ is an internal node given by $\alpha(1+d)^{-\beta}$ where $\alpha\in(0,1)$ and $\beta\in[0,\infty)$ so that $\alpha$ controls how likely a terminal node in the tree would split, with a smaller $\alpha$ implying lesser likelihood a terminal node would split, and $\beta$ controls the number of terminal nodes, and a larger $\beta$ decreasing the number of terminal nodes; (ii) the distribution used to choose which covariate to be selected for the decision rule in an internal node; and (iii) the distribution for the value of the selected covariate for the decision rule in an internal node. Chipman et al. suggests a discrete uniform distribution for the available covariates and values in both (ii) and (iii) respectively, although other more flexible distributions could be used \citep{kapelner_jbart}. 
 
In \cite{chipman_bart}, $\alpha=0.95$ and $\beta=2$. For $\mu_{\mu}$ and $\sigma_{\mu}$, they are set such that $N(m\mu_{\mu},m\sigma_{\mu}^2)$ assigns high probability to the interval $(\underset{k}{\min}(Y_k),\underset{k}{\max}(Y_k))$. This can be achieved by defining $v$ such that $\underset{k}{\min}(Y_k)=m\mu_{\mu}-v\sqrt{m}\sigma_{\mu}$ and $\underset{k}{\max}(Y_k)=m\mu_{\mu}+v\sqrt{m}\sigma_{\mu}$. For convenience when implementing the posterior draws of $T_j$ and $\mu_{ij}$, Chipman et al. suggested transforming the observed $Y_k$ to $\tilde{Y}_k=\frac{Y_k-\frac{\underset{k}{\min}(Y_k)+\underset{k}{\max}(Y_k)}{2}}{\underset{k}{\max}(Y_k)-\underset{k}{\min}(Y_k)}$, and then treating $\tilde{Y}_k$ as the outcome. This has the effect of allowing the hyperparameter of $\mu_{\mu}$ to be set as $\mu_{\mu}=0$ and $\sigma_{\mu}$ to be set as $\sigma_{\mu}=\frac{0.5}{v\sqrt{m}}$ where $v$ is to be chosen. For $v=2$, $N(m\mu_{\mu},m\sigma_{\mu}^2)$ assigns a prior probability of 0.95 to the interval $(\underset{k}{\min}(Y),\underset{k}{\max}(Y))$ and is the suggested value. Finally for $\nu$ and $\lambda$, Chipman et al. suggested setting $\nu=3$ and $\lambda$ is the value such that $P(\sigma^2<s^2;\nu,\lambda)=0.9$ where $s^2$ is the estimated variance of the residuals from the multiple linear regression with $Y_k$ as the outcomes and $\mathbf{X}$ as the covariates.
 
This setup induces the posterior distribution $P[(T_1,\mathbf{M}_1),\ldots,(T_m,\mathbf{M}_m),\sigma|Y_k]$ which can be simplified to two major posterior draws using Gibbs sampling. First, draw $m$ successive 
\begin{equation}
	\label{tree_draws}
	P[(T_j,\mathbf{M}_j)|T_{(j)},\mathbf{M}_{(j)},Y_k,\sigma]
\end{equation}
for $j=1,\ldots,m$, where $T_{(j)}$ and $\mathbf{M}_{(j)}$ consist of all the tree structures and terminal nodes except for the $j^{\text{th}}$ tree structure and terminal node; and then, draw $P[\sigma|(T_1,\mathbf{M}_1),\ldots,(T_m,\mathbf{M}_m),Y_k]$.

To obtain a draw from (\ref{tree_draws}), note that this distribution depends on $(T_{(j)},\mathbf{M}_{(j)},Y_k,\sigma)$ through
\begin{equation}
	\label{gam_meth}
	R_{kj} = Y_k - \sum_{w\neq j}g(\mathbf{X}_k,T_w,\mathbf{M}_w),
\end{equation}
the residuals of the $m-1$ regression sum of trees fit excluding the $j^{\text{th}}$ tree. Thus (\ref{tree_draws}) is equivalent to the posterior draw from a single regression tree $R_{kj}=g(\mathbf{X}_k,T_j,\mathbf{M}_j)+\epsilon_k$ or 
\begin{equation}
	\label{tree_draws_red}
	P[(T_j,\mathbf{M}_j)|R_{kj},\sigma].
\end{equation}
We can obtain a draw from (\ref{tree_draws_red}) by first drawing from $P(T_j|R_{kj},\sigma)$ using a Metropolis-Hastings (MH) algorithm \citep{chipman_bcart, chipman_bart, kapelner_jbart}. A new tree $T_j^*$ can be proposed given the previous tree $T_j$ by four steps: (i) grow, where a terminal node is split into two new child nodes; (ii) prune, where two terminal child nodes immediately under the same non-terminal node is combined together such that their parent non-terminal node becomes a terminal node; (iii) swap, where the splitting criteria of two non-terminal nodes are swapped; (iv) change, where the splitting criteria of a single non-terminal node is changed. Once we draw $P(T_j|R_{kj},\sigma)$, we then draw $P(\mu_{ij}|T_j,R_{kj},\sigma)\sim N(\frac{\sigma_{\mu}^2\sum_i^{n_i}r_{ij}+\sigma^2\mu_{\mu}}{n_i\sigma_{\mu}^2+\sigma^2},\frac{\sigma^2\sigma_{\mu}^2}{n_i\sigma_{\mu}^2+\sigma^2})$, where $r_{ij}$ is the subset of elements in $R_{kj}$ allocated to the terminal node with parameter $\mu_{ij}$ and $n_i$ is the number of $r_{ij}$s in $R_{kj}$ allocated to $\mu_{ij}$. Note that $\mu_{\mu}=0$ after transformation. Complete details for the derivation of $P(\mu_{ij}|T_j,R_{kj},\sigma)$ and $P[\sigma|(T_1,\mathbf{M}_1),\ldots,(T_m,\mathbf{M}_m),Y_k]$ are provided in the supplementary materials available online. Explicit MH algorithm details for equation (\ref{tree_draws_red}) can be found in Kapelner and Bleich.

\subsection{Binary outcomes}
Extending BART to binary outcomes involve a modification of (\ref{bart_main_con}). First, let 
\begin{equation}
	\label{bart_main_bin}
	G(\mathbf{X}_k)=\sum_{j=1}^m g(\mathbf{X}_k,T_j,\mathbf{M}_j).
\end{equation}
Using the probit formulation, the binary outcomes $Y_k$ can be linked to (\ref{bart_main_bin}) using $P(Y_k=1|\mathbf{X}_k)=\Phi[G(\mathbf{X}_k)]$ where $\Phi[.]$ is the cumulative density function of a standard normal distribution. This formulation implicitly assumes that $\sigma\equiv 1$. Assuming once again that all $m$ tree structures and terminal node parameters are independent, this implies that we only need priors for $T_j$ and $\mu_{ij}|T_j$. \cite{chipman_bart} assumes that priors for $T_j$ and $\mu_{ij}$ and the hyperparameters for $\alpha$ and $\beta$ are the same as BART for continuous outcomes. However, for the hyperparameters of $\mu_{\mu}$ and $\sigma_{\mu}$, Chipman et al. suggested that $\mu_{\mu}$ and $\sigma_{\mu}$ should be chosen such that $G(X_k)$ is assigned to the interval $(-3,3)$ with high probability. This can be achieved by setting $\mu_{\mu}=0$ and choosing an appropriate $v$ in the formula $\sigma_{\mu}=\frac{3}{v\sqrt{m}}$. Similar to the continuous outcome case, Chipman et al. suggested $v=2$.

To draw from the posterior distribution $P[(T_1,\mathbf{M}_1),\ldots,(T_m,\mathbf{M}_m)|Y_k]$, \cite{chipman_bart} proposed the use of data augmentation \citep{albert,tanner}. This method proceeds by first generating a latent variable $Z_k$ according to 
\begin{align*}
	(Z_k|Y_k=1,\mathbf{X}_k)&\sim N_{(0,\infty)}(G(\mathbf{X}_k),1)\\
	(Z_k|Y_k=0,\mathbf{X}_k)&\sim N_{(-\infty,0)}(G(\mathbf{X}_k),1),
\end{align*}
where $N_{(a,b)}(\mu,\sigma^2)$ is the truncated normal distribution with mean $\mu$ and variance $\sigma^2$ truncated to the range $(a,b)$. Once $Z_k$ is drawn, $P[(T_1,\mathbf{M}_1),\ldots,(T_m,\mathbf{M}_m)|Z_k]$ is drawn next as in (\ref{tree_draws})-(\ref{tree_draws_red}) with the latent variables $Z_k$ replacing $Y_k$ in (\ref{tree_draws}) and $\sigma$ fixed at 1. Note that at each iteration, $G(\mathbf{X}_k)$ will be updated with the new $(T_1,\mathbf{M}_1),\ldots,(T_m,\mathbf{M}_m)$ draws from $P[(T_1,\mathbf{M}_1),\ldots,(T_m,\mathbf{M}_m)|Z_k]$ so that an updated draw of the latent variable $Z_k$ can be obtained. 

\section{Random Intercept BART}
\subsection{Continuous outcomes}
We now extend BART to account for repeated measurements. We start with the clustered continuous outcomes. We introduce to (\ref{bart_main_con}) a random intercept $a_k$, $k=1,\ldots,K$. Here, $k$ still indexes the subjects but $i=1,\ldots,n_k$ indexes the observations within a subject. With the addition of $a_k$, (\ref{bart_main_con}) becomes
\begin{equation}
	\label{ribart_main_con}
	Y_{ik}=\sum_{j=1}^m g(\mathbf{X}_{ik},T_j,\mathbf{M}_j)+a_k+\epsilon_{ik}\quad\epsilon_{ik}\overset{i.i.d.}{\sim} N(0,\sigma^2),\,a_k\overset{i.i.d.}{\sim} N(0,\tau^2),\,a_k\bot\epsilon_{ik}.
\end{equation}
We decompose the joint prior distribution as (assuming $\sigma^2$ and $\tau^2$ are a priori independent) as
\begin{align*}
	P[(T_1,\mathbf{M}_1),\ldots,(T_m,\mathbf{M}_m),\sigma,\tau]=[\prod_{j=1}^m\{\prod_{l=1}^{b_j}P(\mu_{lj}|T_j)\}P(T_j)]P(\sigma)P(\tau).
\end{align*}
Next, we place the same prior distributions as the independent BART model for $T_j$, $\mu_{lj}|T_j$ (this is $\mu_{ij}$ for the independent BART model), and $\sigma^2$. There are various prior distributions we could place on $\tau^2$ and we discuss this in the next paragraph. We use the same hyperparameter values for $\alpha$, $\beta$, $\mu_{\mu}$, and $\nu$ that \cite{chipman_bart} suggested for the independent BART model. For $\sigma_{\mu}$, we found that $\sigma_{\mu}=\frac{1.8}{v\sqrt{m}}$ worked better for reasons we shall discuss later in this section. For $\lambda$, we first estimated the outcomes $Y_{ik}$ using multiple linear regression (MLR) with $\mathbf{X}_k$ as the predictors. We then estimated an initial random intercept, $\hat{a}^{(0)}_k$, by taking the mean of the MLR residuals for each $k$. Finally, we obtained an initial estimate of $\sigma^2$ using $s^{(0)2}=\frac{\sum_{k=1}^K\sum_{i=1}^{n_k}(Y_{ik}-\hat{Y}^{(0)}_{ik}-\hat{a}^{(0)}_k)^2}{N-p-1-K}$, where $N=\sum_{k=1}^Kn_k$. Then $\lambda$ can be set as the value such that $P(\sigma^2<s^{(0)2};\nu,\lambda)=0.9$. We call this model the random intercept BART (riBART). 

To test the sensitivity of riBART to different prior distributions of $\tau^2$, we tried first, a flat improper prior, $P(\tau^2)\propto 1$; second, a half-Cauchy prior \citep{gelman}, achieved by reformulating the random intercept as
\begin{equation}
	\label{riBART_decom_b}
	a_k = \xi\eta_k \quad \xi\sim N(0,B^2),\,\eta_k\overset{i.i.d.}{\sim} N(0,\theta^2)
\end{equation}
and assuming that $B^2$, $\theta^2$, $\sigma^2$ and $(T_j,\mathbf{M}_j)$s are independent, $\theta^2\sim IG(\frac{1}{2},\frac{1}{2})$ and $B=25$; and finally, a proper prior, $\tau^2\sim IG(1,1)$. For the half-Cauchy prior, the posterior draws of $a_k$ and $\tau$ can be obtained by setting $a_k=\xi\eta_k$ and $\tau=|\xi|\theta$.

To draw from the posterior distribution of riBART, we employ a Metropolis within Gibbs procedure. We first draw the Gibbs sample of $\sigma$, $\tau$, and $a_k$ separately from their respective posterior distribution. Then, using the updated $a_k$, we obtain $\tilde{Y}_{ik}=Y_{ik}-a_k$. Now $\tilde{Y}_{ik}|\mathbf{X}_k$ can be viewed as a BART model. The idea of viewing $\tilde{Y}_{ik}|\mathbf{X}_k$ as a BART model has been discussed in \cite{zhang} and \cite{dorie}. To allow convenient implementation of the posterior draws of $T_j$ and $\mu_{lj}|T_j$, we transform the outcomes $\tilde{Y}_{ik}$ to $\check{Y}_{ik}=\frac{3.6[\tilde{Y}_{ik}-\frac{\underset{i,k}{\min}(\tilde{Y}_{ik})+\underset{i,k}{\max}(\tilde{Y}_{ik})}{2}]}{\underset{i,k}{\max}(\tilde{Y}_{ik})-\underset{i,k}{\min}(\tilde{Y}_{ik})}$. This transformation produced posterior draws for $\sigma$ and $\tau$ with better repeated sampling properties across the range of our simulation studies compared to the usual transformation employed in BART, and suggests setting $\sigma_{\mu}=\frac{1.8}{2\sqrt{m}}$ so that $(\underset{i,k}{\min}(\tilde{Y}_{ik}),\underset{i,k}{\max}(\tilde{Y}_{ik}))$ has a prior probability of 0.95. We suspect this transformation produces better repeated sampling properties for the posterior draws of $\sigma$ and $\tau$ because it allows $\check{Y}_{ik}$ to vary more. Further investigation beyond the scope of this paper is needed in order to determine why this is the case. After obtaining $\check{Y}_{ik}$, we use $\check{Y}_{ik}$ as the outcome in the BART algorithm to obtain the posterior distribution of $T_j$. In our implementation, we employed the grow and prune steps for the proposal of a new tree $T_j^*$ for computational ease. Given $T_j$, we then draw $\mu_{lj}$. Derivation of the Gibbs sampling distributions of $\sigma$, $a_k$, $\tau$, $\theta^2$, and $\eta_k$ are provided in the supplementary materials available online.

\subsection{Binary outcomes}
Extending riBART to binary outcomes proceed in a similar fashion. We add $a_k$ to (\ref{bart_main_bin}) to obtain
\begin{equation}
	\label{ribart_main_bin}
	G_a(\mathbf{X}_{ik})=\sum_{j=1}^m g(\mathbf{X}_{ik},T_j,\mathbf{M}_j)+a_k.
\end{equation} 
We once again assume $a_k\sim N(0,\tau^2)$. To link the sum of trees to the binary outcomes $Y_{ik}$, we use the probit link and write $P(Y_{ik}=1|\mathbf{X}_{ik})=\Phi[G_a(\mathbf{X}_{ik})]$. We suggest prior distributions similar to the continuous outcomes riBART for $T_j$, $\mu_{lj}$, and $\tau^2$. The same hyperparameters in BART for binary outcome can be used for $\alpha$, $\beta$, $\mu_{\mu}$, and $\sigma_{\mu}$. To obtain the posterior draws of $T_j$, $\mathbf{M}_j$, $a_k$, and $\tau^2$, we employ the data augmentation method suggested by \cite{albert96}. First, we draw a latent variable $Z_{ik}$ according to 
\begin{align*}
	(Z_{ik}|Y_{ik}=1,\mathbf{X}_{ik})&\sim N_{(0,\infty)}(G_a(\mathbf{X}_{ik}),1)\\
	(Z_{ik}|Y_{ik}=0,\mathbf{X}_{ik})&\sim N_{(-\infty,0)}(G_a(X_{ik}),1).
\end{align*}
We then draw $\tau$ followed by $a_k$. Next, we remove $a_k$ from $Z_{ik}$ to obtain $\tilde{Z}_{ik}=Z_{ik}-a_k$. $\tilde{Z}_{ik}|\mathbf{X}_{ik}$ can now be viewed as a continuous BART model and the usual BART algorithm can be applied with $\sigma$ fixed at 1. In our implementation, we employed a further transformation of $\tilde{Z}_{ik}$ to $\check{Z}_{ik}=\frac{6[\tilde{Z}_{ik}-\frac{\underset{i,k}{\min}(\tilde{Z}_{ik})+\underset{i,k}{\max}(\tilde{Z}_{ik})}{2}]}{\underset{i,k}{\max}(\tilde{Z}_{ik})-\underset{i,k}{\min}(\tilde{Z}_{ik})}$. This keeps $\check{Z}_{ik}$ within the range of $(-3,3)$, which we found produces posterior draws for $\tau$ with better repeated sampling properties across the range of our simulation studies. The posterior draw is then completed by updating $Z_{ik}$ using the most recent posterior draws of $(T_1,\mathbf{M}_1),\ldots,(T_m,\mathbf{M}_m)$, and $a_k$.    

\section{Simulation Study}
We conducted a simulation study to determine the bias, root mean squared error (RMSE), 95\% coverage, and average 95\% credible interval length (AIL) of riBART compared to BART on a longitudinal dataset with correlated outcomes. The models we compared were: (I) BART, (II) riBART with $P(\tau^2)\propto 1$ (flat), (III) riBART with half-Cauchy prior on $\tau^2$ (half-Cauchy), and (IV) riBART with $\tau^2\sim IG(1,1)$ (proper). The parameters we focused on were $\sum_{j=1}^m g(\mathbf{X}_{ik},T_j,\mathbf{M}_j)+a_k$ abbreviated as $g(x)+a_k$, $\sigma$, and $\tau$. We also investigated the MSE (continuous) and AUC (binary) produced by each model.  

We generated our correlated outcomes dataset by first drawing the predictors using $X_{ikq}\overset{i.i.d.}{\sim}\text{Uniform}(0,1)$, $q=1,\ldots,5$. For continuous outcomes, we generated 
\begin{equation}
	\label{sim_dat_con}
	Y_{ik}=10\sin(\pi X_{ik1}X_{ik2})+20(X_{ik3}-0.5)^2+10X_{ik4}+5X_{ik5}+a_k+\epsilon_{ik}
\end{equation}
where $\epsilon_{ik}\overset{i.i.d.}{\sim} N(0,\sigma^2)$ and $a_k\overset{i.i.d.}{\sim} N(0,\tau^2)$. For binary outcomes, we first generated  
\begin{equation}
	\label{sim_dat_bin}
	G_a(X_{ik})=1.35[\sin(\pi X_{ik1}X_{ik2})+2(X_{ik3}-0.5)^2-X_{ik4}-0.5X_{ik5}]+a_k
\end{equation}
where $a_k\overset{i.i.d.}{\sim} N(0,\tau^2)$. Then, we generated the binary outcomes $Y_{ik}$ by drawing $Z_{ik}\sim N(G_a(\mathbf{X}_{ik}),1)$ and setting $Y_{ik}=1$ if $Z_{ik}>0$, otherwise $Y_{ik}=0$. 

For the study design, we considered $K=50$ clusters with $n_k=5$ observations per cluster and $K=100$ clusters with $n_k=20$ observations per cluster. We also considered $\tau=0.5$ and $\tau=1$. This produces eight different simulation scenarios summarized in Tables \ref{biascover_con} and \ref{biascover_bin}. For each simulation, we conducted 1,000 burn ins followed by 5,000 posterior draws. Bias, RMSE, 95\% coverage, AIL, MSE, and AUC were estimated from 200 simulations for each scenario. All our simulations were done in \textit{R 3.1.1} \citep{rprog}.

Table \ref{biascover_con} shows the bias, RMSE, 95\% coverage and AIL of $\sum_{j=1}^m g(\mathbf{X}_{ik},T_j,\mathbf{M}_j)+a_k$, $\sigma$, and $\tau$ under continuous correlated outcomes. We observed that the bias, RMSE, and 95\% coverage for $\sum_{j=1}^m g(\mathbf{X}_{ik},T_j,\mathbf{M}_j)+a_k$ were similar and reasonable for both BART and riBART models with BART having a tendency to under cover $\sum_{j=1}^m g(\mathbf{X}_{ik},T_j,\mathbf{M}_j)+a_k$ when sample size increases. In addition, riBART produces a wider 95\% credible interval on average because it takes into consideration the additional $a_k$ parameter. For the bias, RMSE, 95\% coverage, and AIL of $\sigma$, BART tended to produce more absolute bias and poorer coverage. On average, RMSE was smaller for riBART methods. Although the AIL were similar, riBART clearly produced better 95\% coverage for $\sigma$ except when $n_k=20$ and $K=100$ where 95\% coverage of $\sigma$ for riBART methods were around 83-85\%, about 10\% less than the nominal rate. We believe this is caused by the regression trees getting stuck at certain tree structures in the MH algorithm and hence variation of the $\sigma$ parameter is affected. We shall discuss this further in Section \ref{discussion}. For $\tau$, the half-Cauchy prior did not seem to work well when $n_k=5$ and $K=50$ in terms of bias, RMSE, and 95\% coverage. The $\tau^2\sim IG(1,1)$ prior worked the best in terms of bias, RMSE, 95\% coverage, and AIL when $n_k=5$, $K=50$, and $\tau=0.5$ while the $P(\tau^2)\propto 1$ prior worked slightly better when $n_k=5$, $K=50$, and $\tau=1$. When $n_k=20$ and $K=100$, all three priors produced similar results for the estimation of $\tau$ in terms of bias, RMSE, 95\% coverage, and AIL.

For binary correlated outcomes, the main focus of our paper, we found that bias, RMSE, and 95\% coverage of $\sum_{j=1}^m g(\mathbf{X}_{ik},T_j,\mathbf{M}_j)+a_k$ were often poorer for BART except for the bias of $\sum_{j=1}^m g(\mathbf{X}_{ik},T_j,\mathbf{M}_j)+a_k$ under $n_k=20$, $K=100$, and $\tau=0.5$, where the bias of $\sum_{j=1}^m g(\mathbf{X}_{ik},T_j,\mathbf{M}_j)+a_k$ in BART was smaller compared to all three riBART methods. Similar to the continuous correlated outcomes, the AIL for BART was smaller compared to riBART mainly because BART ignores the estimation of the parameter, $a_k$. For the bias, RMSE, 95\% coverage, and AIL of $\tau$, all three riBART methods produced similar results except when $n_k=5$, $K=50$, and $\tau=0.5$, where the riBART under $\tau^2\sim IG(1,1)$ produced more bias, RMSE, and lower 95\% coverage for $\tau$.

\begin{figure}[H]
	\caption{Boxplots of mean squared error (MSE) for continuous correlated outcomes produced by BART, riBART with $P(\tau^2)\propto 1$, half-Cauchy prior on $\tau^2$, and $\tau^2\sim IG(1,1)$. \label{simMSE}}
	\hspace*{-1.25cm}
	\centering
	\begin{tabular}{cc}
		(a) $n_k=5$, $K=50$, $\tau=1$, $\sigma=1$  & (b) $n_k=20$, $K=100$, $\tau=1$, $\sigma=1$ \\
		\includegraphics[scale=0.5]{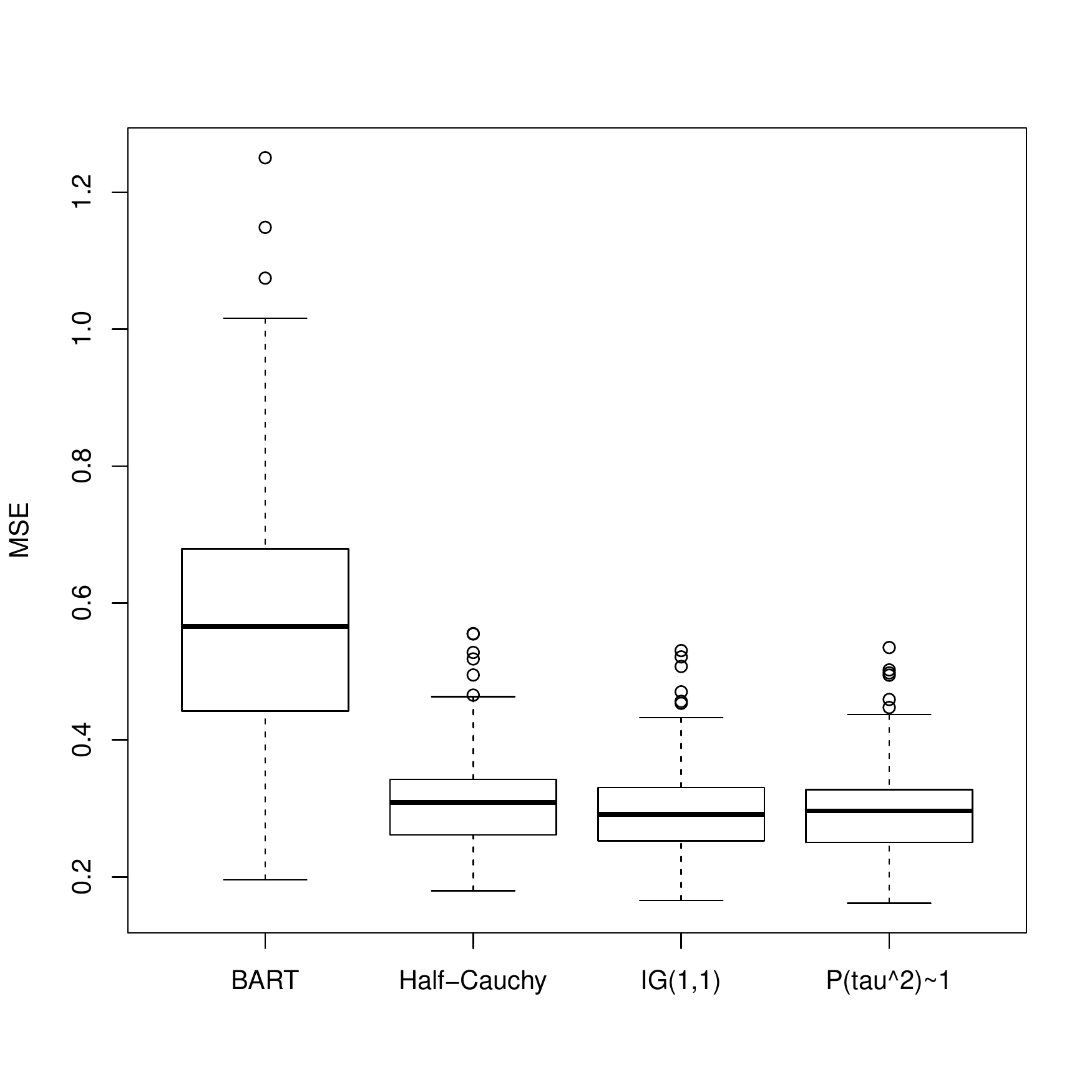} &  \includegraphics[scale=0.5]{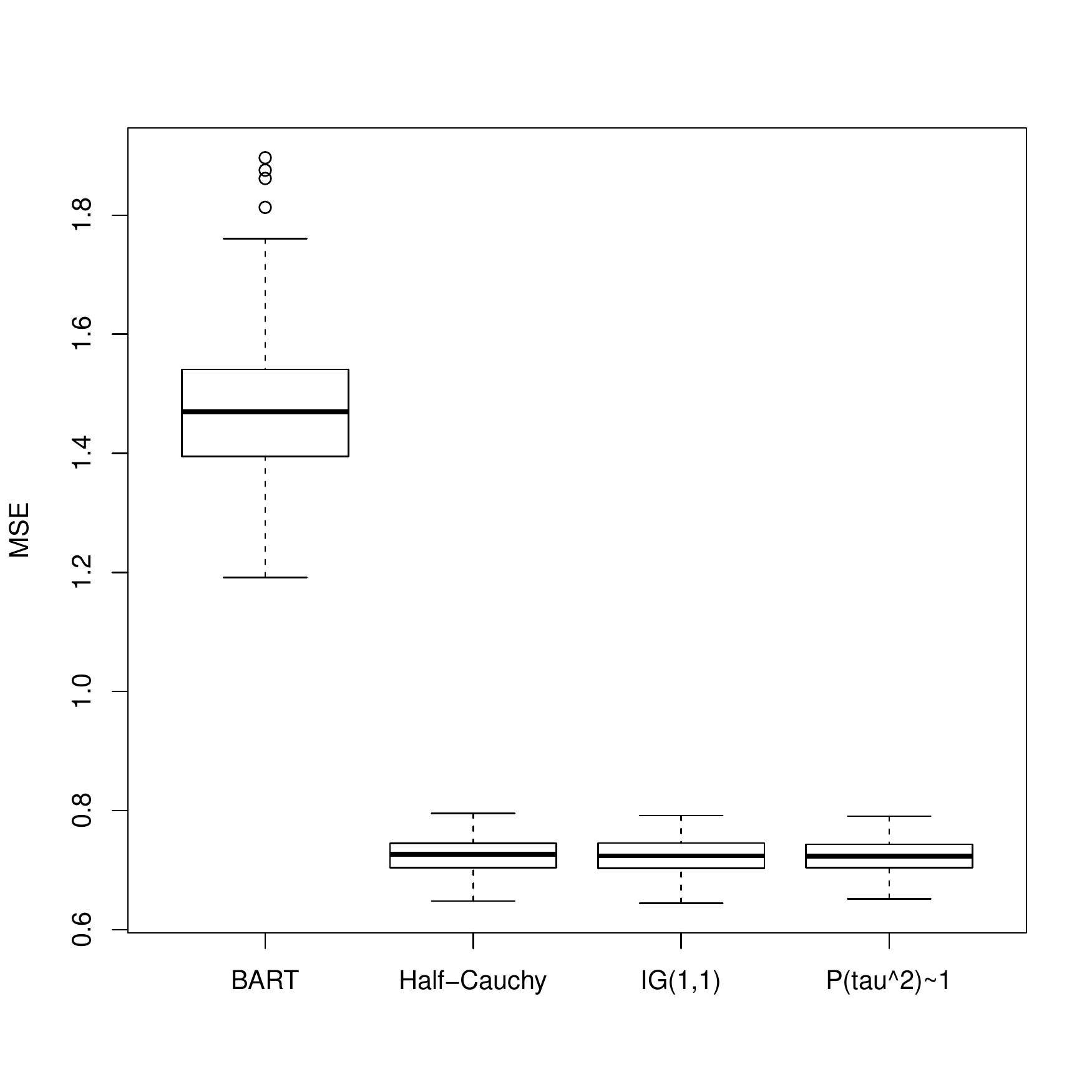} \\
		(c) $n_k=5$, $K=50$, $\tau=0.5$, $\sigma=1$  & (d) $n_k=20$, $K=100$, $\tau=0.5$, $\sigma=1$ \\
	 	\includegraphics[scale=0.5]{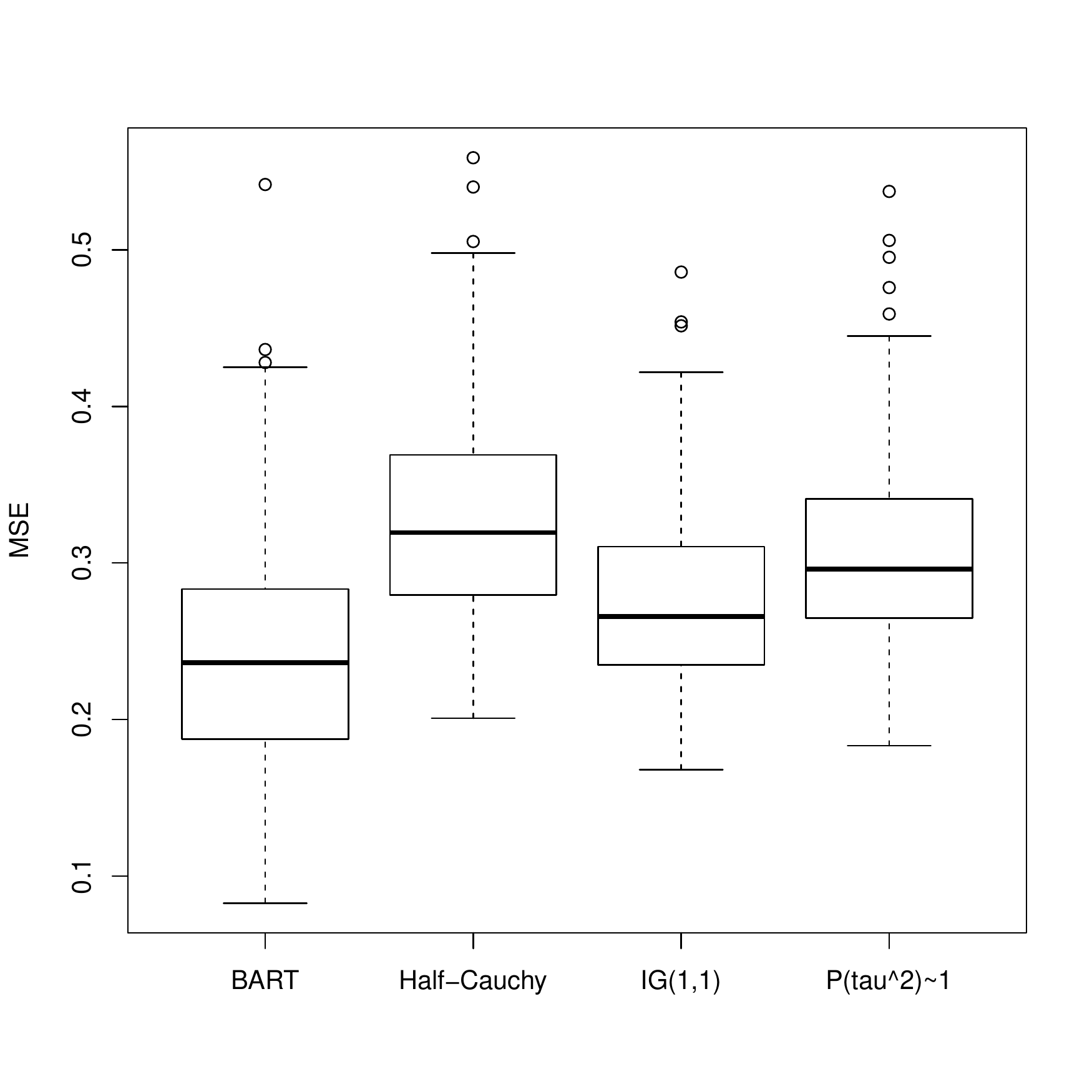} &  \includegraphics[scale=0.5]{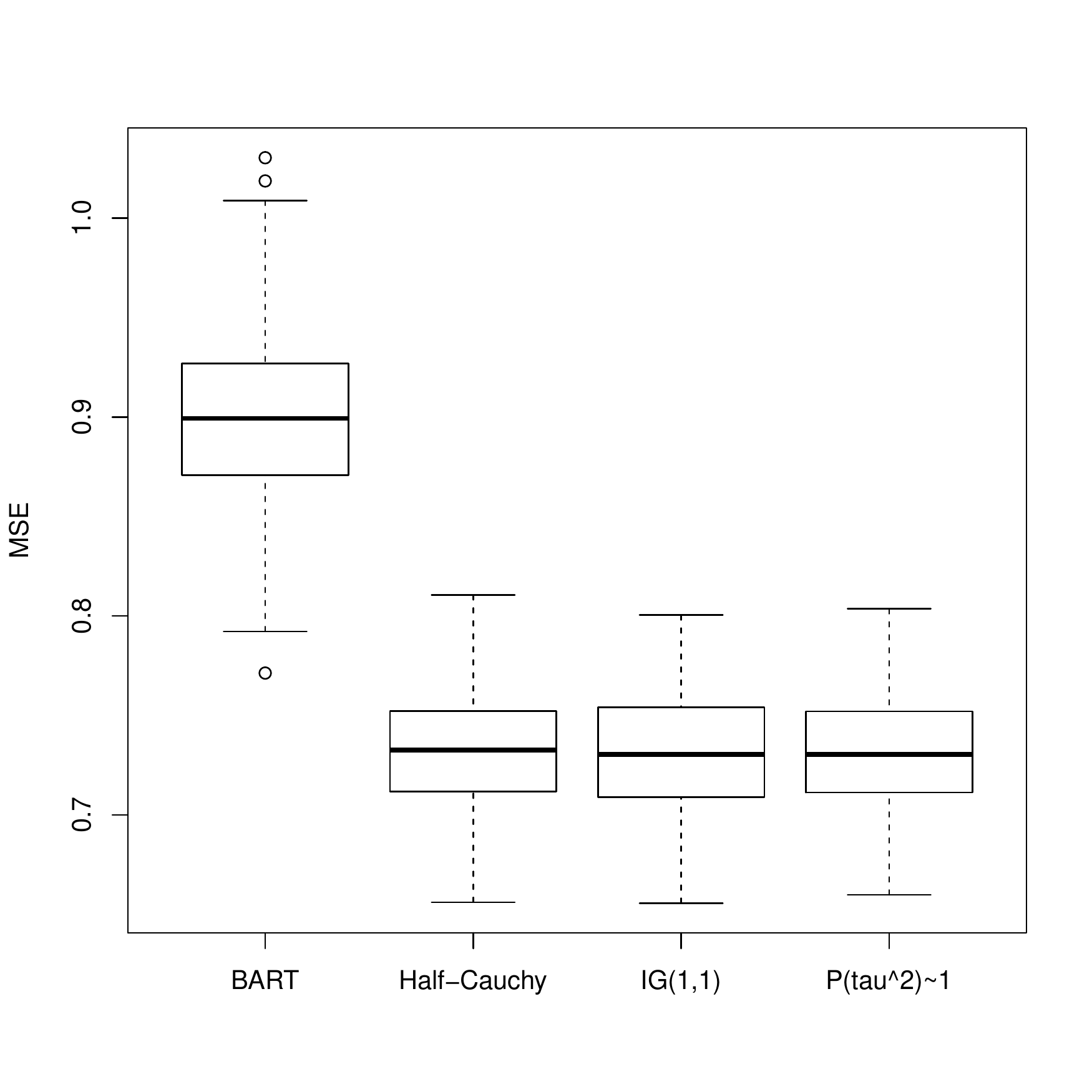} \\
	\end{tabular}
\end{figure}

\begin{figure}[H]
	\caption{Boxplots of area under the receiver operating characteristic curve (AUC) for binary correlated outcomes produced by BART, riBART with $P(\tau^2)\propto 1$, half-Cauchy prior on $\tau^2$, and $\tau^2\sim IG(1,1)$. \label{simAUC}}
	\hspace*{-1.25cm}
	\centering
	\begin{tabular}{cc}
		(a) $n_k=5$, $K=50$, $\tau=1$  & (b) $n_k=20$, $K=100$, $\tau=1$ \\
		\includegraphics[scale=0.5]{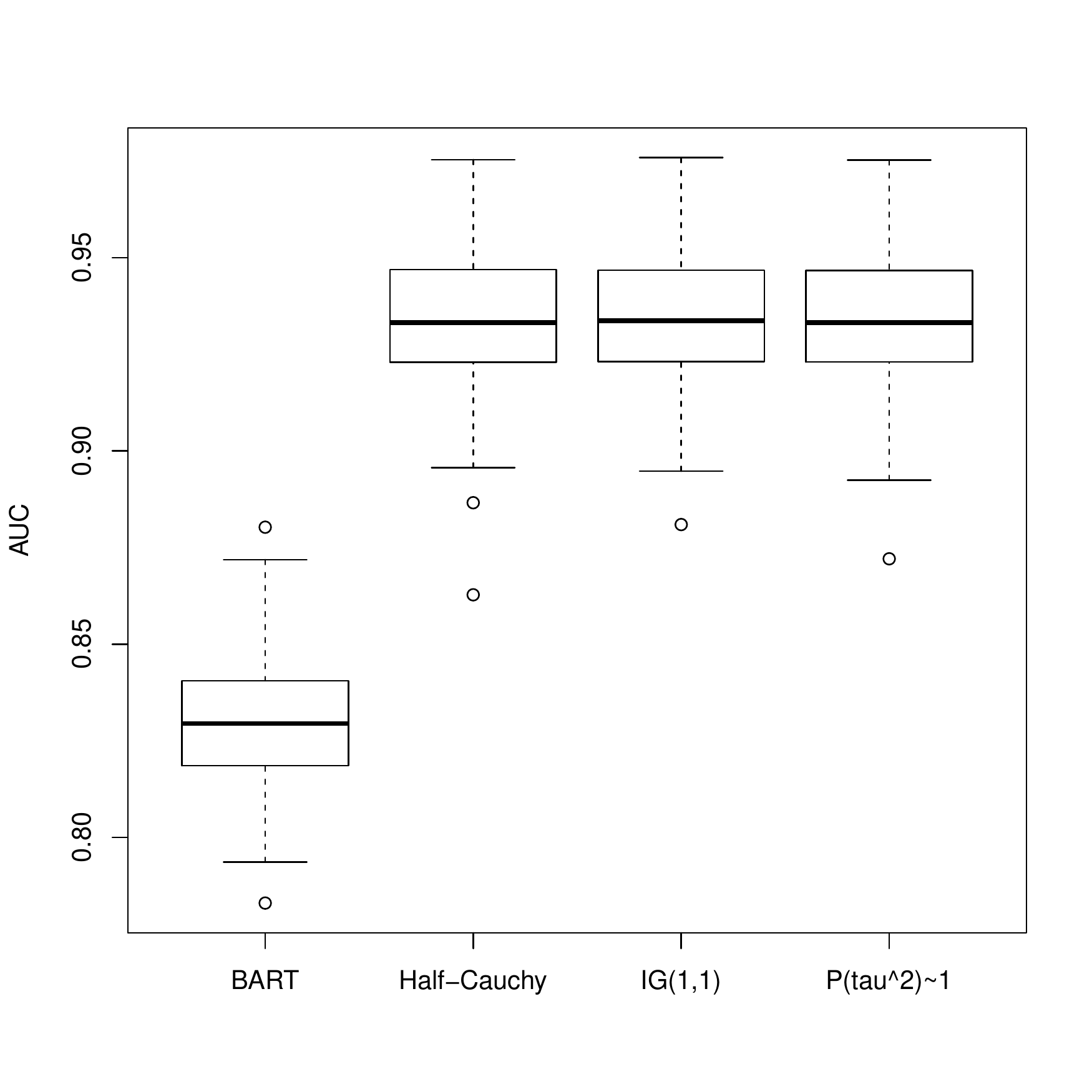} & \includegraphics[scale=0.5]{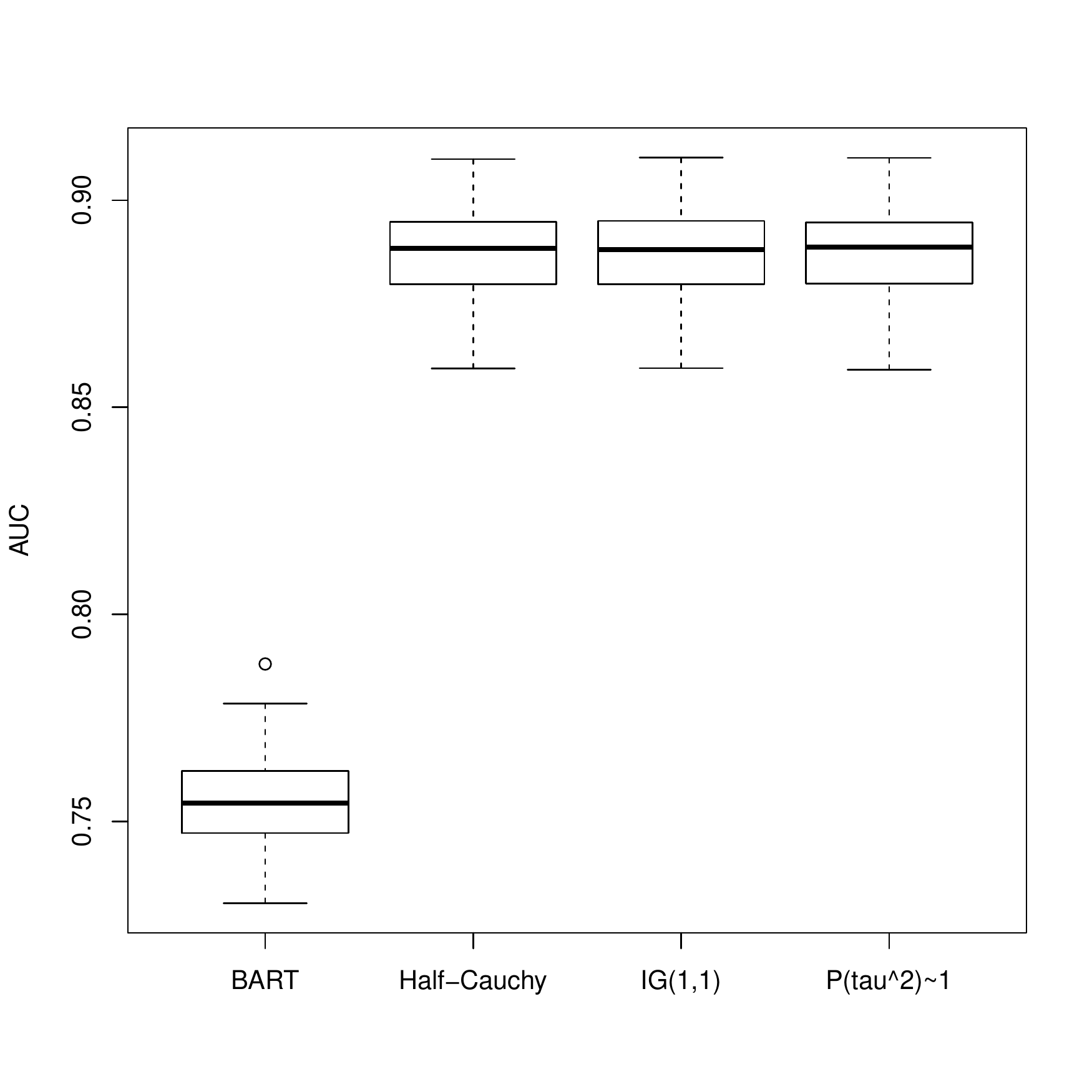} \\
		(c) $n_k=5$, $K=50$, $\tau=0.5$  & (d) $n_k=20$, $K=100$, $\tau=0.5$ \\
	 	\includegraphics[scale=0.5]{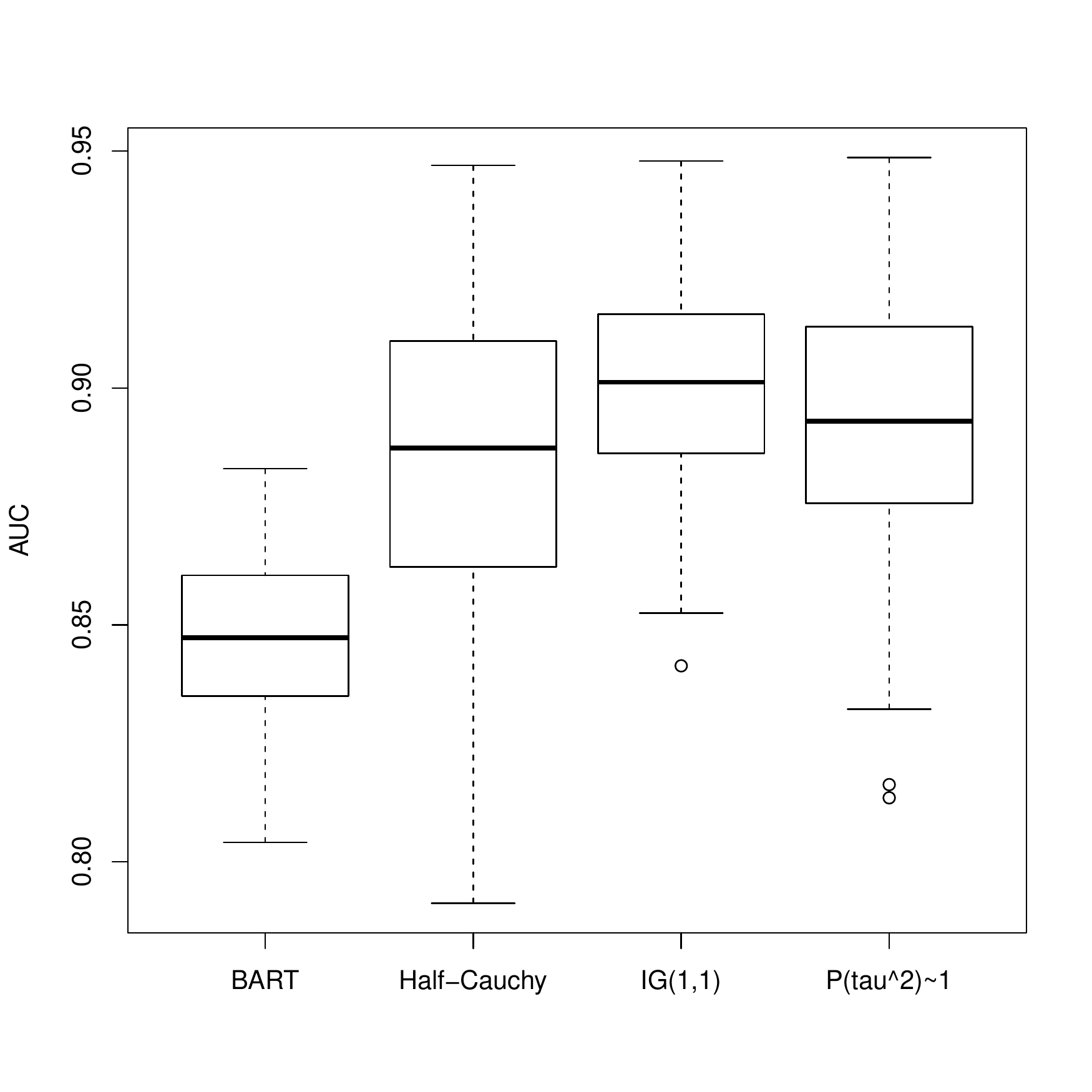} & \includegraphics[scale=0.5]{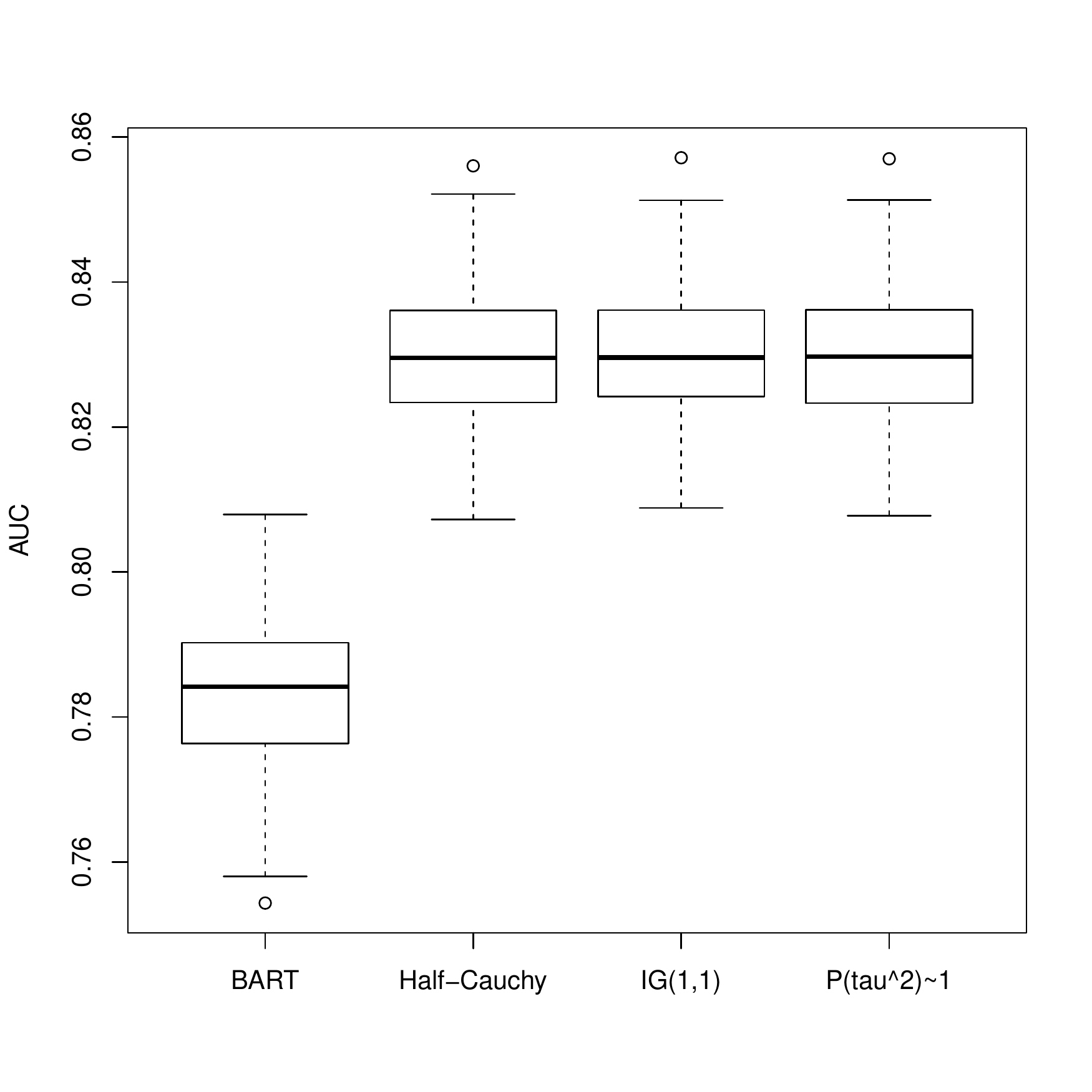} \\
	\end{tabular}
\end{figure}

Figure \ref{simMSE} shows the boxplots of the MSEs for scenarios 1 to 4 while Figure \ref{simAUC} shows the boxplots of the AUCs produced for scenarios 5 to 8. Other than the setting where $n_k=5$ and $\tau=0.5$, the MSEs of continuous outcomes riBART under the three $\tau^2$ prior distributions were all smaller compared to BART. In addition, there does not seem to be much difference in the MSE between riBART under the three different $\tau^2$ prior distributions. For binary correlated outcomes, we again observed that AUC for riBART was higher compared to BART for all correlated binary outcomes scenarios except when $n_k=5$ and $\tau=0.5$. Again AUC produced by riBART under the three different $\tau^2$ priors were similar.

In summary, the 3 different prior distributions on $\tau^2$ for riBART does not seem to produce be much difference in the estimation of $\sum_{j=1}^m g(\mathbf{X}_{ik},T_j,\mathbf{M}_j)+a_k$ in terms of the bias, RMSE, 95\% coverage, and AIL. In addition, MSE and AUC were rather similar for all 3 riBART methods. For continuous correlated outcomes with $\tau=1$, riBART with $P(\tau^2)\propto 1$ is preferred because it produces better repeated sampling properties for the posterior draws of $\sigma$ and $\tau$. For $\tau=0.5$, we prefer riBART with $\tau^2\sim IG(1,1)$ because of similar reasons. For binary correlated outcomes, we prefer riBART with half-Cauchy prior on $\tau^2$ for $\tau=1$ and for $\tau=0.5$, when $n_k=5$ and $K=50$
because better repeated sampling properties for the posterior draws of $\tau$ were produced. When $n_k=20$, $K=1000$, and $\tau=0.5$, riBART with $\tau^2\sim IG(1,1)$ is preferred instead because of similar reasons. 

\section{Predicting Driver Stop before Left Turn Execution} 
\subsection{Integrated Vehicle-Based Safety Systems (IVBSS) Study}
The dataset we used to develop our prediction model was obtained from the Integrated Vehicle Based Safety System (IVBSS) study conducted by \cite{sayer}. This study collected naturalistic driving data from 108 licensed drivers in Michigan between April 2009 and April 2010. In the study, sixteen late-model Honda Accords were fitted with cameras, recording devices, and several integrated collision warning systems. Each driver used a vehicle for a total of 40 days -- 12 days baseline period with IVBSS switched off followed by 28 days with IVBSS activated. Since our objective was to develop a prediction model for human driving behavior, we used the 12 days baseline unsupervised driving data. In total the 108 drivers made 3,795 turns, of which 1,823 were left turns. Each driver took on average of 35 turns, with a range of 8 to 139 turns per driver. This suggests that riBART could potentially improve the prediction performance of our model compared to BART, while simultaneously accounting for the correlation among observations in inference.

\subsection{Analysis}
To begin prediction, we extracted both the speed of the vehicle (in m/s) and the distance traveled (in m) at 10 millisecond intervals starting from 100 meters away from the center of an intersection. To obtain a practical prediction model, we converted the time series of vehicle speeds to a distance series to provide a distance-varying definition for our binary outcomes of whether a vehicle would stop before executing a left turn in the future. Our outcome was whether a vehicle would eventually stop before executing a left turn, estimated repeatedly at 1 meter intervals before the intersection. For example, if the vehicle's current location is -45 meters, the outcome is whether the vehicle will stop between -44 and -1 meter. If a vehicle stops and restarts, the outcome is reset: a vehicle that stops at -40 meters and then proceeds through the intersection will have an outcome of 1 (stopping) from -94 to -40 meters, and 0 (not stopping) from -39 to -1 meters. 

At any given distance, we could use the full profile of a vehicle's past speeds as the predictors, but these speeds may contain irrelevant information. Thus, we determined a moving window of recent speeds to be used in our prediction model at every meter. Using a 10-fold cross validation with AUC as the judging criteria and BART as the model, we selected an optimal window length of 6 meters. To further reduce the number of variables to consider in our model, we then used Principal Components Analysis (PCA) to summarize the vehicle speeds in each 6 meter moving window. The first three PC scores explained more than 99\% of the variation in the vehicle speed and we found that adding PC scores beyond these did not improve prediction. We included a fourth predictor, the number of times the vehicle has stopped up to the current location. This fourth predictor adjusts for the likely correlation within each turn. 

We conducted a preliminary data analysis using logistic regression, BART, and a Super Learner ensemble method \citep{vanderLaan} that considered  elastic net \citep{friedman_enet}, logistic regression, K-Nearest Neighbor, Generalized Additive Models \citep{hastie_book}, mean of the outcomes, and BART. Super Learner and BART had similar prediction performance as measured by AUC, but BART was far more stable.

We fit riBART with a random effect at the driver level which incorporates within-driver correlation. Based on our simulation results, we used the proper prior ($IG(1,1)$) for $\tau^2$. For comparison, we also ran BART, which ignores within-driver correlation; and a random intercept linear logistic regression, which incorporates within-driver correlation but ignores non-linearity and complex interactions. For these models, we used the same distance-varying predictors and outcome as riBART. The linear logistic regression was obtained using the \textit{glm} function in \textit{R} while the random intercept linear logistic regression was obtained using the \textit{glmer} function from the \textit{R} package \textit{lme4}. We also computed the 95\% CI of the AUCs using the method of \cite{hanley}, which uses a linear approximation of the AUC to the Somer's D statistic to obtain an estimate of the variance of AUC.

\subsection{Results}
\begin{figure}[H]
	\caption{(a) The intra-class correlation (ICC) profile of riBART as a factor of distance from the intersection; (b) Area under the receiver operating characteristic curve (AUC) profile  of riBART, BART, and random intercept logistic regression (dotted lines are 95\% Credible Interval); and (c) AUC difference profile between riBART versus BART and riBART versus random intercept linear logistic regression. \label{anal_res}}
	\hspace*{-1.25cm}
	\centering
	\begin{tabular}{cc}
	(a) ICC & (b) AUC \\
	\includegraphics[scale=0.5]{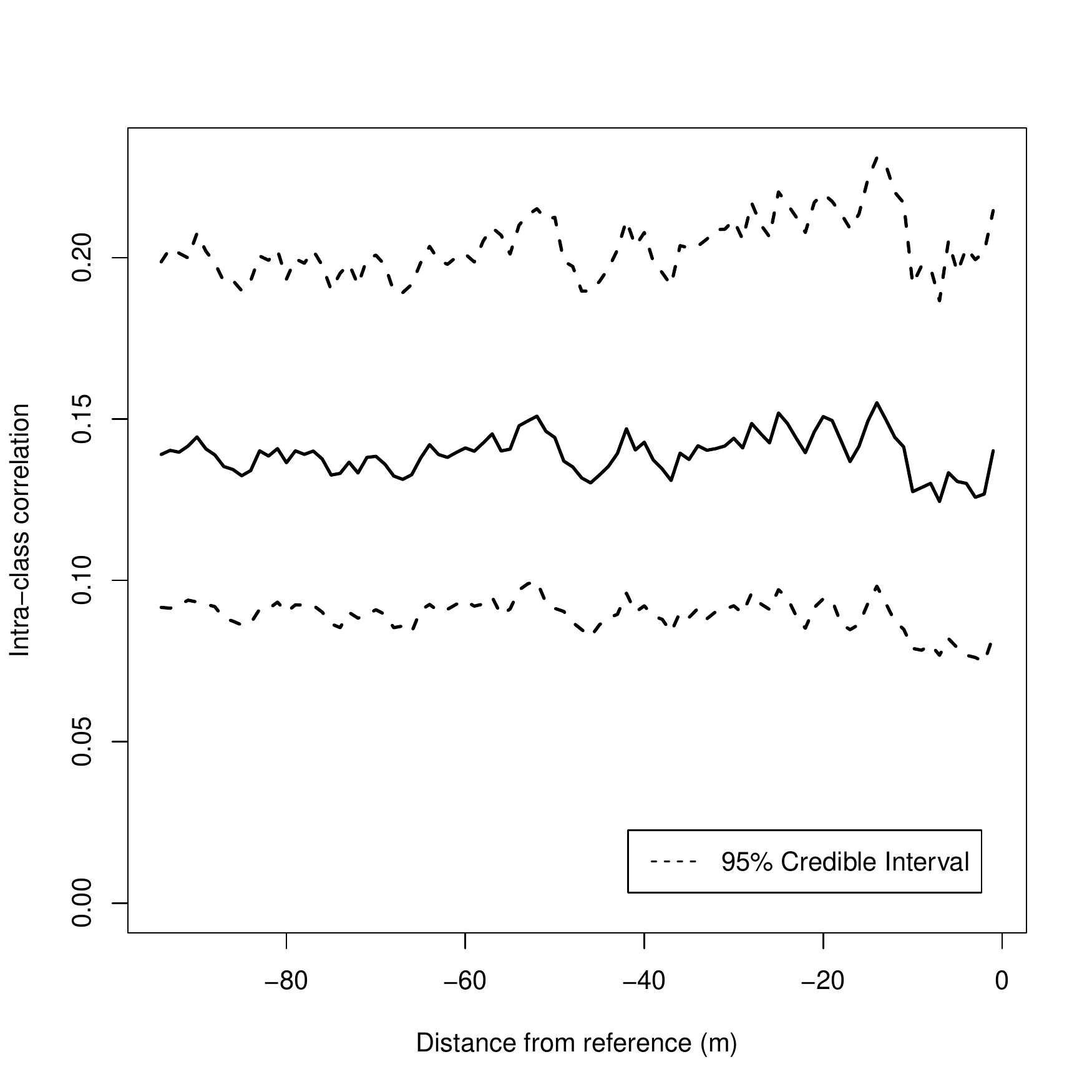} & \includegraphics[scale=0.5]{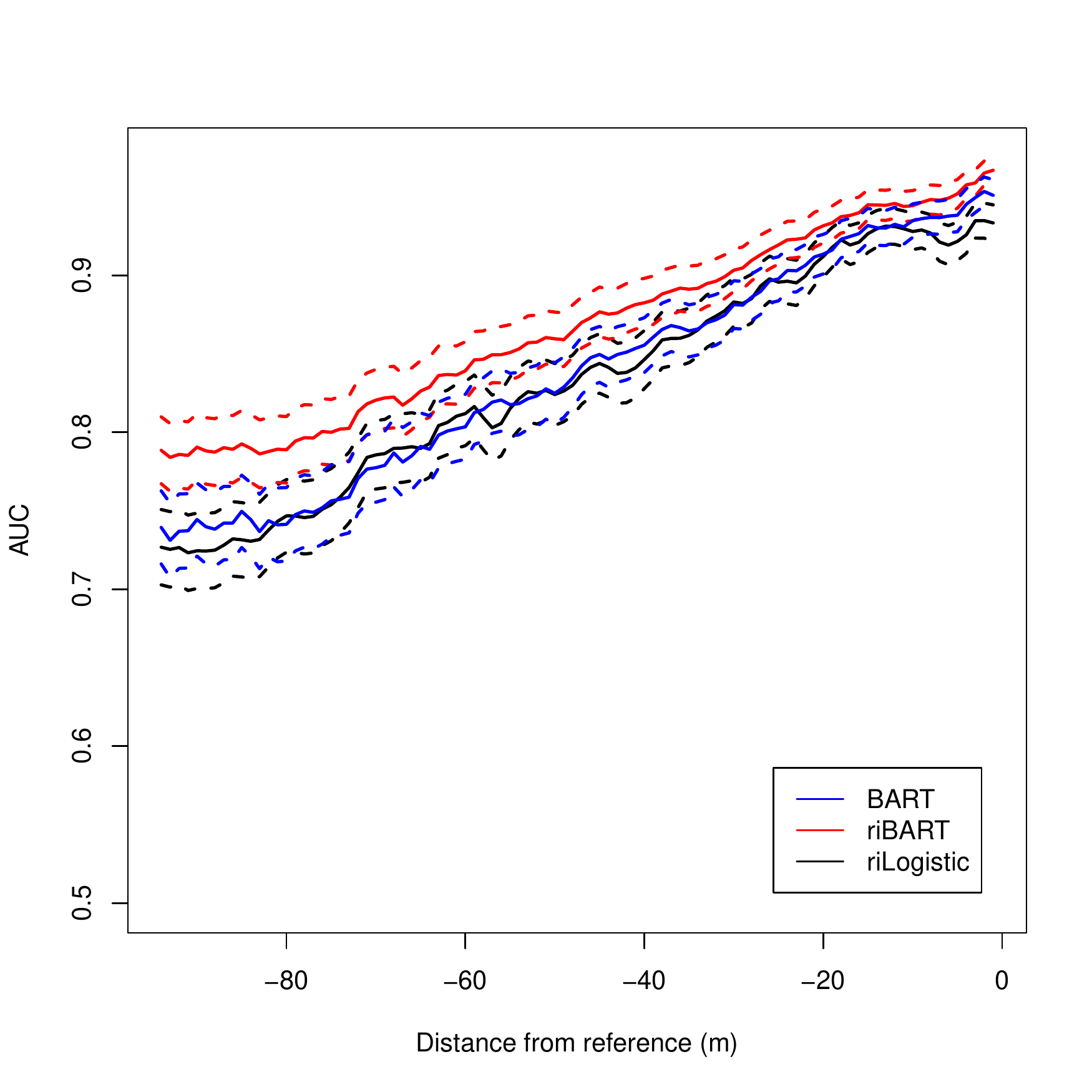} \\
	(c) AUC difference versus riBART \\
	\includegraphics[scale=0.5]{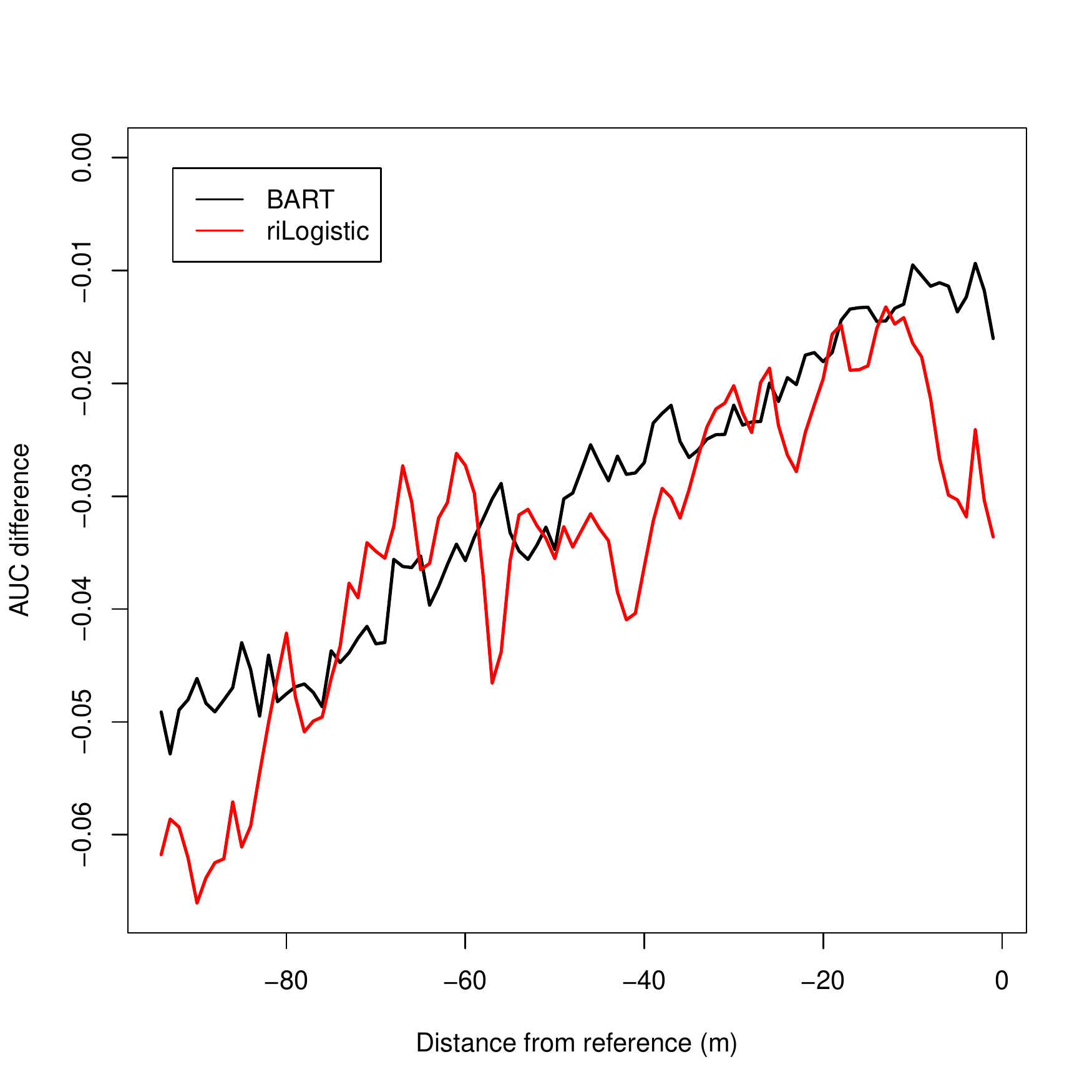}
	\end{tabular}
\end{figure}

Figure \ref{anal_res} shows (a) the the estimated intra-class correlation (ICC, $\frac{\tau^2}{\tau^2+1}$) profile; (b) the AUC profiles of riBART, BART, and random intercept linear logistic regression; and (c) the AUC profile difference between riBART versus BART, and riBART versus random intercept linear logistic regression.

The posterior mean profile of ICC was small, between about 0.12 and 0.16, and fairly stable as the vehicle approaches the center of an intersection. This suggests firstly that the variance parameter, $\tau$, for the random intercept, $a_k$, is small for left turn stops and secondly that as the vehicle approaches the center of the intersection, the effect of individual `habits' of the driver remained relatively stable throughout the left turn maneuver. For the AUC profile (middle), we see evidence that riBART performs better than both BART and random intercept linear logistic regression. Both BART and random intercept linear logistic regression perform similarly in terms of AUC. BART produced an AUC estimate of about 0.74 with an estimated 95\% CI of (0.72, 0.76) at -94m away from the center of intersection. For both riBART specifications, the AUC was about 0.78 [95\% C.I. (0.76, 0.80)] at -94m away from the center of intersection. The difference in AUC profile between riBART versus BART and riBART versus random intercept linear logistic regression remained negative throughout the left turn maneuver with the absolute difference decreasing as the vehicle approaches the center of an intersection. 

\section{Discussion \label{discussion}}
In this paper, we developed a model, riBART, to help engineers developing self driving vehicles predict whether a human-driven vehicle would stop at an intersection before executing a left turn. We achieved this by utilizing the model that did well in our preliminary analyses, BART, and extending it to account for the key feature in our dataset, clustered observations. Although existing methods extending BART to longitudinal datasets were available, our approach was more straight-forward and can be implemented on correlated binary outcomes. Codes implementing riBART can be made available upon request. Applying riBART to our dataset, substantial improvement in prediction compared to BART can be obtained when we take into account that different drivers have different `propensities to stop' before executing a left turn at an intersection; that is, the inclusion of a random intercept improves prediction performance for our dataset compared to a model without a random intercept. This implies that future development of an operational algorithm should try to accommodate the similarities of stopping behavior for a given human driver through a learning algorithm. For example, devices that are able to transmit information about a driver's propensity to stop could be installed on vehicles to improve the decision-making performance of the self driving vehicle.

In our simulation study, we found that the true 95\% coverage for a 95\% posterior prediction interval for $\sigma$ was reduced when the number of clusters and the number of observations within a cluster was large ($n_k=20$, $K=100$). The likely cause for the poor coverage is due to low variation in the posterior draw of $\sigma$ resulting in reduced average 95\% credible interval length. We believe this low variation in $\sigma$ is due to the regression trees in BART getting stuck at certain tree structures. This phenomenon of regression trees getting stuck at certain tree structures has been discussed by \cite{pratola} previously. The difference here is that Pratola only reported observing regression trees being stuck when the true $\sigma$ is small for regression trees. We argue that regression trees might also get stuck when the effective sample size, $N$, is large because with a large $N$, deeper trees would needed to get a better fit of $R_{kj}$ in equation (\ref{gam_meth}). When a regression tree gets deep, the standard grow, prune, change, and swap steps may have trouble proposing new trees with radically different tree structures. This lack of radically different tree structures implies reduced variability in the tree structures, which is indirectly reflected by the lack of variation in $\sigma$. 

This issue is separate from the development of BART in the correlated data context, and indeed would occur when observations are independent. We illustrate this with an example using BART implemented via the \textit{BayesTree} package in \textit{R}. We generated $Y_k=10\sin(\pi X_{k1}X_{k2})+20(X_{k3}-0.5)^2+10X_{k4}+5X_{k5}+\epsilon_k$ with $X_{kq}\overset{i.i.d.}{\sim}\text{Uniform}(0,1)$, $q=1,\ldots,5$ and  $\epsilon_{ik}\overset{i.i.d.}{\sim} N(0,1)$. We then ran 200 simulations each with a different signal function but keeping $\sigma=1$ for all simulations. The sample size we used in all 200 simulations was 2,000. The resulting bias, RMSE, 95\% coverage, and AIL for $\sigma$ was -0.04, 0.04, 79\%, and 0.09 respectively. We observe once again that although bias and RMSE were small, the 95\% coverage for $\sigma$ was far from nominal because the AIL was small. We think that this issue of a lack in variation of $\sigma$ when the sample size is large could be solved by either increasing the number of regression trees used, re-calibrating the $\alpha$ and $\beta$ parameters used to penalize each regression tree, or to include the rotate step proposed by \cite{pratola} in the proposal of a new regression tree in the MH algorithm of BART. As inference about $\sigma$ is not the key focus of this paper, we leave investigation of this problem with BART to future work.

Our proposed model only included a random intercept but, there may be situations where the researcher believes that there may be more complicated linear random effect mechanisms occurring in the real world setting. In our application, estimating a ``turn-level'' random effect nested within the driver-level random effect could have been done but would be of little value for predicting future turns. However, in other settings, estimating and splitting of these variance components might be useful. Other plausible areas for future research include extending BART and riBART to outcomes of other forms, for example, ordinal outcomes or counts.

\section*{Acknowledgments}
This work was supported jointly by Dr. Michael Elliott and an ATLAS Research Excellence Program project awarded to Dr. Carol Flannagan. This work was also funded in part by the Toyota Class
Action Settlement Safety Research and Education Program. The conclusions are those of the authors and have not been sponsored, approved, or endorsed by Toyota or plaintiffs' class counsel. We would like to thank Dr. Jian Kang and Dr. Brisa S\'{a}nchez for their valuable suggestions.

\section*{Supplementary materials}
\subsection*{Posterior distributions for $\mu_{ij}$ and $\sigma^2$ in BART}
\subsubsection*{$P(\mu_{ij}|T_j,\sigma,R_{ij})\sim N(\frac{\sigma_{\mu}^2\sum_i^{n_i}r_{ij}+\sigma^2\mu_{\mu}}{n_i\sigma_{\mu}^2+\sigma^2},\frac{\sigma^2\sigma_{\mu}^2}{n_i\sigma_{\mu}^2+\sigma^2})$:}

Let $R_{ij}=(r_{1j},\ldots, r_{n_ij})$ be a subset from $R_{kj}$ where $n_i$ is the number of $r_{ij}$s allocated to the terminal node with parameter $\mu_{ij}$. We note that $R_{ij}|g(X_{ik},T_j,M_j),\sigma\sim N(\mu_{ij},\sigma^2)$ and $\mu_{ij}|T_j\sim N(\mu_{\mu},\sigma_{\mu}^2)$. Then the posterior distribution of $\mu_{ij}$ is given by
\begin{align*}
	P(\mu_{ij}|T_j,\sigma,R_{ij})&\propto P(R_{ij}|T_j,\mu_{ij},\sigma)P(\mu_{ij}|T_j)\\
	&\propto \exp[-\frac{\sum_i(r_{ij}-\mu_{ij})^2}{2\sigma^2}]\exp[-\frac{(\mu_{ij}-\mu_{\mu})^2}{2\sigma_{\mu}^2}]\\
	&\propto \exp[-\frac{(n_i\sigma_{\mu}^2+\sigma^2)\mu_{ij}^2-2(\sigma_{\mu}^2\sum_ir_{ij}+\sigma^2\mu_{\mu})\mu_{ij}}{2\sigma^2\sigma_{\mu}^2}]\\
	&\propto\exp[-\frac{(\mu_{ij}-\frac{\sigma_{\mu}^2\sum_ir_{ij}+\sigma^2\mu_{\mu}}{n_i\sigma_{\mu}^2+\sigma^2})^2}{2\frac{\sigma^2\sigma_{\mu}^2}{n_i\sigma_{\mu}^2+\sigma^2}}]
\end{align*}
where $\sum_i(r_{ij}-\mu_{ij})^2$ is the summation of the squared difference between the parameter $\mu_{ij}$ and the $r_{ij}$s allocated to the terminal node with parameter $\mu_{ij}$. 

\subsubsection*{$P(\sigma|(T_1,M_1),\ldots,(T_m,M_m),Y)\sim IG(\frac{\nu+n}{2},\frac{\nu\lambda+\sum_{k=1}^n(y_k-\sum_{j=1}^m g_k(X_k,T_j,M_j))^2}{2})$:}
Let $Y=(y_1,\ldots,y_n)$ and $k$ index the subjects $k=1,\ldots,n$. With $\sigma^2\sim IG(\frac{\nu}{2},\frac{\nu\lambda}{2})$, we obtain the posterior draw of $\sigma$ as follows
\begin{align*}
	P(\sigma|(T_1,M_1),\ldots,(T_m,M_m),Y)&\propto P(Y|(T_1,M_1),\ldots,(T_m,M_m),\sigma)P(\sigma^2)\\
	&= P(Y|\sum_{j=1}^m g(X_k,T_j,M_j),\sigma)P(\sigma^2)\\
	&=\{\prod_{k=1}^n(\sigma^2)^{-\frac{1}{2}}\exp[-\frac{(y_k-\sum_{j=1}^m g_k(X_k,T_j,M_j))^2}{2\sigma^2}]\}\\
	&\quad(\sigma^2)^{-(\frac{\nu}{2}+1)}\exp(-\frac{\nu\lambda}{2\sigma^2})\\
	&=(\sigma^2)^{-(\frac{\nu+n}{2}+1)}\exp[-\frac{\nu\lambda+\sum_{k=1}^n(y_k-\sum_{j=1}^m g_k(X_k,T_j,M_j))^2}{2\sigma^2}]
\end{align*}
where $\sum_j^m g_k(X_k,T_j,M_j)$ is the predicted value of BART assigned to observed outcome $y_k$. 

\subsection*{Posterior distributions of $a_k$ and $\sigma^2$ for riBART}

In this section, $k$ still indexes the subjects and while $i$ now indexes the number of repeated measures for each subject i.e. $i=1,\ldots,n_k$. Let $Y=(y_{11},\ldots,y_{1n_1},\ldots,y_{K1},\ldots,y_{Kn_K})$ and $\hat{y}_{ik}=\sum_{j=1}^m g(X_{ik},T_j,M_j)$.

\subsubsection*{$P(a_k|Y,(T_1,M_1),\ldots,(T_m,M_m),\sigma,\tau)\sim N(\frac{\tau^2\sum_{i=1}^{n_k}(y_{ik}-\hat{y}_{ik})}{n_k\tau^2+\sigma^2},\frac{\sigma^2\tau^2}{n_k\tau^2+\sigma^2})$:}

Since $a_k\sim N(0,\tau^2)$, we have
\begin{align*}
	P(a_k|Y,(T_1,M_1),\ldots,(T_m,M_m),\sigma,\tau)&\propto P(Y|\sum_{j=1}^m g(X_{ik},T_j,M_j),\sigma,a_k)P(a_k|\tau^2)\\
	&\propto \{\prod_{i=1}^{n_k}\exp[-\frac{(y_{ik}-\hat{y}_{ik}-a_k)^2}{2\sigma^2}]\}\exp[-\frac{a_k^2}{2\tau^2}]\\
	&\propto \exp[-\frac{\sum_{i=1}^{n_k}(y_{ik}-\hat{y}_{ik}-a_k)^2}{2\sigma^2}]\exp[-\frac{a_k^2}{2\tau^2}]\\
	&\propto \exp[-\frac{(n_k\tau^2+\sigma^2)a_k^2-2\tau^2a_k\sum_{i=1}^{n_k}(y_{ik}-\hat{y}_{ik})}{2\sigma^2\tau^2}]\\
	&=\exp[-\frac{(a_k-\frac{\tau^2\sum_{i=1}^{n_k}(y_{ik}-\hat{y}_{ik})}{n_k\tau^2+\sigma^2})^2}{2\frac{\sigma^2\tau^2}{n_k\tau^2+\sigma^2}}].
\end{align*}

\subsubsection*{$P(\sigma^2|Y,(T_1,M_1),\ldots,(T_m,M_m),a_k,\tau)\sim IG(\frac{N+\nu}{2},\frac{\sum_{k=1}^K\sum_{i=1}^{n_k}(y_{ik}-\hat{y}_{ik}-a_k)^2+\nu\lambda}{2})$:}

For the posterior of $\sigma^2$, since we have $\sigma^2\sim IG(\frac{\nu}{2},\frac{\nu\lambda}{2})$, we obtain
\begin{align*}
	P(\sigma^2|Y,(T_1,M_1),\ldots,(T_m,M_m),a_k,\tau)&\propto P(Y|\sum_{j=1}^m g(X_{ik},T_j,M_j),\sigma,a_k)P(\sigma^2)\\
	&\propto\{\prod_{k=1}^K\prod_{i=1}^{n_k}(\sigma^2)^{-\frac{1}{2}}\exp[-\frac{(y_{ik}-\hat{y}_{ik}-a_k)^2}{2\sigma^2}]\}\\
	&\quad(\sigma^2)^{-(\frac{\nu}{2}+1)}\exp[-\frac{\nu\lambda}{2\sigma^2}]\\
	&\propto (\sigma^2)^{-(\frac{N+\nu}{2}+1)}\\
	&\quad\exp[-\frac{\sum_{k=1}^K\sum_{i=1}^{n_k}(y_{ik}-\hat{y}_{ik}-a_k)^2+\nu\lambda}{2\sigma^2}]
\end{align*}
where $\sum_{k=1}^Kn_k=N$.

\subsection*{Posterior distribution of $\tau$ under $P(\tau^2)\propto 1$ and $\tau^2\sim IG(1,1)$}

\subsubsection*{$\tau^2|Y,(T_1,M_1),\ldots,(T_m,M_m),a_k,\sigma\sim IG(\frac{K}{2}-1,\frac{\sum_{k=1}^Ka_k^2}{2})$ for $P(\tau^2)\propto 1$:}
 
\begin{align*}
	P(\tau^2|Y,(T_1,M_1),\ldots,(T_m,M_m),a_k,\sigma)&\propto \{\prod_{k=1}^K P(a_k|\tau^2)\}P(\tau)\\
	&\propto (\tau^2)^{-\frac{K}{2}}\exp[-\frac{\sum_{k=1}^Ka_k^2}{2\tau^2}].
\end{align*} 

\subsubsection*{$\tau^2|Y,(T_1,M_1),\ldots,(T_m,M_m),a_k,\sigma\sim IG(\frac{K}{2}+1,\frac{\sum_{k=1}^Ka_k^2+2}{2})$ for $\tau^2\sim IG(1,1)$:}

\begin{align*}
	P(\tau^2|Y,(T_1,M_1),\ldots,(T_m,M_m),a_k,\sigma)&\propto \{\prod_{k=1}^K P(a_k|\tau^2)\}P(\tau)\\
	&\propto (\tau^2)^{-\frac{K}{2}}\exp[-\frac{\sum_{k=1}^Ka_k^2}{2\tau^2}](\tau^2)^{-(1+1)}\exp[-\frac{1}{\tau^2}]\\
	&\propto (\tau^2)^{-(\frac{K}{2}+1+1)}\exp[-\frac{\sum_{k=1}^Ka_k^2+2}{2\tau^2}].
\end{align*}

\subsection*{Posterior distributions for $\xi$, $\eta_k$, $\theta$ and $\sigma^2$ for riBART with half-Cauchy prior on $\tau^2$}

\subsubsection*{$P(\xi|Y,(T_1,M_1),\ldots,(T_m,M_m),\eta_k,\theta,\sigma)\sim N(\frac{\sum_{k=1}^K\sum_{i=1}^{n_k}\eta_k(y_{ik}-\hat{y}_{ik})}{\sum_{k=1}^K\sum_{i=1}^{n_k}\eta_k^2},\frac{\sigma^2}{\sum_{k=1}^K\sum_{i=1}^{n_k}\eta_k^2})$:}
We note that $\xi\sim N(0,B^2)$, $\eta_k\sim N(0,\theta^2)$, $\sigma^2\sim \nu\lambda\chi_{\nu}^2$, and $\theta^2\sim ef/\chi_f^2$. Now for
\begin{align*}
	P(\xi|Y,(T_1,M_1),\ldots,(T_m,M_m),\eta_k,\theta,\sigma)&\propto P(Y|\sum_{j=1}^m g(X_{ik},T_j,M_j),\sigma,\eta_k,\xi)P(\xi)\\
	&\propto \{\prod_{k=1}^K\prod_{i=1}^{n_k}(\sigma^2)^{-\frac{1}{2}}\exp[-\frac{(y_{ik}-\hat{y}_{ik}-\xi\eta_k)^2}{2\sigma^2}]\}\\
	&\quad\exp[-\frac{\xi^2}{2B^2}]\\
	&\propto \exp[-\frac{(\xi-\frac{B^2\sum_{k=1}^K\sum_{i=1}^{n_k}\eta_k(y_{ik}-\hat{y}_{ik})}{B^2\sum_{k=1}^K\sum_{i=1}^{n_k}\eta_k^2+\sigma^2})^2}{2\frac{\sigma^2B^2}{B^2\sum_{k=1}^K\sum_{i=1}^{n_k}\eta_k^2+\sigma^2}}].
\end{align*}
is the kernel of a $N(\frac{B^2\sum_{k=1}^K\sum_{i=1}^{n_k}\eta_k(y_{ik}-\hat{y}_{ik})}{B^2\sum_{k=1}^K\sum_{i=1}^{n_k}\eta_k^2+\sigma^2},\frac{\sigma^2B^2}{B^2\sum_{k=1}^K\sum_{i=1}^{n_k}\eta_k^2+\sigma^2})$. Applying l'Hopital's rule taking $B\rightarrow\infty$ yields $N(\frac{\sum_{k=1}^K\sum_{i=1}^{n_k}\eta_k(y_{ik}-\hat{y}_{ik})}{\sum_{k=1}^K\sum_{i=1}^{n_k}\eta_k^2},\frac{\sigma^2}{\sum_{k=1}^K\sum_{i=1}^{n_k}\eta_k^2})$.

\subsubsection*{$P(\eta_k|Y,(T_1,M_1),\ldots,(T_m,M_m),\xi,\theta,\sigma)\sim N(\frac{\theta^2\xi\sum_{i=1}^{n_k}(y_{ik}-\hat{y}_{ik})}{\theta^2\xi^2n_k+\sigma^2},\frac{\sigma^2\theta^2}{\theta^2\xi^2n_k+\sigma^2})$:}

\begin{align*}
	P(\eta_k|Y,(T_1,M_1),\ldots,(T_m,M_m),\xi,\theta,\sigma)&\propto P(Y|\sum_{j=1}^m g(X_{ik},T_j,M_j),\sigma,\eta_k,\xi)P(\eta_k)\\
	&\propto \{\prod_{i=1}^{n_k}(\sigma^2)^{-\frac{1}{2}}\exp[-\frac{(y_{ik}-\hat{y}_{ik}-\xi\eta_k)^2}{2\sigma^2}]\}\\
	&\quad \exp[-\frac{\eta_k^2}{2\theta^2}]\\
	&\propto \exp[-\frac{(\eta_k-\frac{\theta^2\xi\sum_{i=1}^{n_k}(y_{ik}-\hat{y}_{ik})}{\theta^2\xi^2n_k+\sigma^2})^2}{2\frac{\sigma^2\theta^2}{\theta^2\xi^2n_k+\sigma^2}}].
\end{align*}

\subsubsection*{$P(\theta^2|Y,(T_1,M_1),\ldots,(T_m,M_m),\xi,\eta_k,\sigma)\sim IG(\frac{e+K}{2},\frac{\sum_{k=1}^K\eta_k^2+ef}{2}$):}

\begin{align*}
	P(\theta^2|Y,(T_1,M_1),\ldots,(T_m,M_m),\xi,\eta_k,\sigma)&\propto \{\prod_{k=1}^K p(\eta_k|\theta^2)\}p(\theta^2)\\
	&\propto (\theta^2)^{-\frac{K}{2}}\exp[-\frac{\sum_{k=1}^K\eta_k^2}{2\theta^2}] (\theta^2)^{-(\frac{e}{2}-1)}\exp[-\frac{ef}{2\theta^2}]\\
	&\propto (\theta^2)^{-(\frac{e+K}{2}-1)}\exp[-\frac{\sum_{k=1}^K\eta_k^2+ef}{2\theta^2}].
\end{align*}

\subsubsection*{$P(\sigma^2|Y,(T_1,M_1),\ldots,(T_m,M_m),\xi,\eta_k,\theta)\sim IG(\frac{N+\nu}{2},\frac{\sum_{k=1}^K\sum_{i=1}^{n_k}(y_{ik}-\hat{y}_{ik}-\xi\eta_k)^2+\nu\lambda}{2})$:}

\begin{align*}
	P(\sigma^2|Y,(T_1,M_1),\ldots,(T_m,M_m),\xi,\eta_k,\theta)&\propto P(Y|\sum_{j=1}^m g(X_{ik},T_j,M_j),\sigma,\xi,\eta_k,\theta)P(\sigma^2)\\
	&\propto\{\prod_{k=1}^K\prod_{i=1}^{n_k}(\sigma^2)^{-\frac{1}{2}}\exp[-\frac{(y_{ik}-\hat{y}_{ik}-\xi\eta_k)^2}{2\sigma^2}]\}\\
	&\quad(\sigma^2)^{-(\frac{\nu}{2}+1)}\exp[-\frac{\nu\lambda}{2\sigma^2}]\\
	&\propto (\sigma^2)^{-(\frac{N+\nu}{2}+1)}\\
	&\quad\exp[-\frac{\sum_{k=1}^K\sum_{i=1}^{n_k}(y_{ik}-\hat{y}_{ik}-\xi\eta_k)^2+\nu\lambda}{2\sigma^2}]
\end{align*}

\bibliography{references}

\begin{thebibliography}{}

\bibitem[\protect\citeauthoryear{Albert and Chib}{Albert and
  Chib}{1993}]{albert}
Albert, J. and Chib, S. ({1993}).
\newblock {Bayesian Analysis of Binary and Polychotomous Response Data}.
\newblock {\em {Journal of the American Statistical Association}} {\bf {88},}
  {669--679}.

\bibitem[\protect\citeauthoryear{Albert and Chib}{Albert and
  Chib}{1996}]{albert96}
Albert, J. and Chib, S. ({1996}).
\newblock {\em {Bayesian modeling of binary repeated measures data with
  application to crossover trials}}.
\newblock In Bayesian Biostatistics, D. A. Berry and D. K. Stangl, eds. New
  York: Marcel Dekker.

\bibitem[\protect\citeauthoryear{Breiman, Friedman, Olshen, and Stone}{Breiman
  et~al.}{1984}]{breiman}
Breiman, L., Friedman, J., Olshen, R., and Stone, C. ({1984}).
\newblock {\em {Classification and regression Trees}}.
\newblock Wadsworth, Belmont, CA.

\bibitem[\protect\citeauthoryear{Chipman, George, and McCulloch}{Chipman
  et~al.}{1998}]{chipman_bcart}
Chipman, H., George, E., and McCulloch, R. ({1998}).
\newblock {Bayesian CART Model Search}.
\newblock {\em {Journal of the American Statistical Association}} {\bf {93},}
  {935--948}.

\bibitem[\protect\citeauthoryear{Chipman, George, and McCulloch}{Chipman
  et~al.}{2010}]{chipman_bart}
Chipman, H., George, E., and McCulloch, R. ({2010}).
\newblock {BART: Bayesian Additive Regression Trees}.
\newblock {\em {The Annals of Applied Statistics}} {\bf {4},} {266--298}.

\bibitem[\protect\citeauthoryear{{Davies, A.}}{{Davies, A.}}{2015}]{gm}
{Davies, A.} (2015).
\newblock {GM Has `Aggressive' Plans for Self-Driving Cars, Retrieved May 15,
  2016, from
  \verb?https://www.wired.com/2015/10/gm-has-aggressive-plans-for-self-driving-cars/?}

\bibitem[\protect\citeauthoryear{Dorie, Harada, Carnegie, and Hill}{Dorie
  et~al.}{2016}]{dorie}
Dorie, V., Harada, M., Carnegie, N., and Hill, J. ({2016}).
\newblock {A flexible, interpretable framework for assessing sensitivity to
  unmeasured confounding}.
\newblock {\em {Statistics in Medicine}} page {doi:10.1002/sim.6973}.

\bibitem[\protect\citeauthoryear{Franke}{Franke}{1982}]{franke}
Franke, R. ({1982}).
\newblock {Smooth interpolation of scattered data by local thin plate splines}.
\newblock {\em {Computers and Mathematics with Applications}} {\bf {8},}
  {273--281}.

\bibitem[\protect\citeauthoryear{Friedman, Hastie, and Tibshirani}{Friedman
  et~al.}{2010}]{friedman_enet}
Friedman, J., Hastie, T., and Tibshirani, R. ({2010}).
\newblock {Regularization Paths for Generalized Linear Models via Coordinate
  Descent}.
\newblock {\em {Journal of Statistical Software}} {\bf {33},} {1--22}.

\bibitem[\protect\citeauthoryear{Gammermann}{Gammermann}{2000}]{gammermann}
Gammermann, A. ({2000}).
\newblock {Support vector machine learning algorithm and transduction}.
\newblock {\em {Computational Statistics}} {\bf {5},} {31--39}.

\bibitem[\protect\citeauthoryear{Gelman}{Gelman}{2006}]{gelman}
Gelman, A. ({2006}).
\newblock {Prior distributions for variance parameters in hierarchical models
  (Comment on Article by Browne and Draper)}.
\newblock {\em {Bayesian Analysis}} {\bf {1},} {515--534}.

\bibitem[\protect\citeauthoryear{{Google}}{{Google}}{2015}]{google}
{Google} (2015).
\newblock {What we’re up to, Retrieved August 26, 2015, from
  \verb?http://www.google.com/selfdrivingcar/?}

\bibitem[\protect\citeauthoryear{Hanley and McNeil}{Hanley and
  McNeil}{1982}]{hanley}
Hanley, J. and McNeil, B. ({1982}).
\newblock {The Meaning and Use of the Area under a Receiver Operating
  Characteristic (ROC) Curve}.
\newblock {\em {Radiology}} {\bf {143},} {29--36}.

\bibitem[\protect\citeauthoryear{Hastie and Tibshirani}{Hastie and
  Tibshirani}{1990}]{hastie_book}
Hastie, T. and Tibshirani, R. ({1990}).
\newblock {\em {Generalized additive models}}.
\newblock CRC Press: Boca Raton, FL.

\bibitem[\protect\citeauthoryear{Kapelner and Bleich}{Kapelner and
  Bleich}{2016}]{kapelner_jbart}
Kapelner, A. and Bleich, J. ({2016}).
\newblock {bartMachine: Machine Learning with Bayesian Additive Regression
  Trees}.
\newblock {\em {Journal of Statistical Software}} {\bf {70},} {1--40}.

\bibitem[\protect\citeauthoryear{Low-Kam, Telesca, Ji, Zhang, Xia, Zink, and
  Nel}{Low-Kam et~al.}{2015}]{lowkam}
Low-Kam, C., Telesca, D., Ji, Z., Zhang, H., Xia, T., Zink, J., and Nel, A.
  ({2015}).
\newblock {A Bayesian regression tree approach to identify the effect of
  nanoparticles' properties on toxicity profiles}.
\newblock {\em {The Annals of Applied Statistics}} {\bf {9},} {383--401}.

\bibitem[\protect\citeauthoryear{{Mchugh, M.}}{{Mchugh, M.}}{2015}]{tesla}
{Mchugh, M.} (2015).
\newblock {Tesla’s Cars Now Drive Themselves, Kinda, Retrieved May 15, 2016,
  from
  \verb?http://www.wired.com/2015/10/tesla-self-driving-over-air-update-live/?}

\bibitem[\protect\citeauthoryear{Pratola}{Pratola}{2016}]{pratola}
Pratola, M. ({2016}).
\newblock {Efficient Metropolis-Hastings Proposal Mechanisms for Bayesian
  Regression Tree Models}.
\newblock {\em {Bayesian Analysis}} {\bf {11},} {885--911}.

\bibitem[\protect\citeauthoryear{{R Core Team}}{{R Core Team}}{2015}]{rprog}
{R Core Team} (2015).
\newblock {\em R: A Language and Environment for Statistical Computing}.
\newblock R Foundation for Statistical Computing, Vienna, Austria.

\bibitem[\protect\citeauthoryear{R\"{a}ssler}{R\"{a}ssler}{2002}]{rassler}
R\"{a}ssler, S. ({2002}).
\newblock {\em {Statistical matching: A frequentist theory, practical
  applications and alternative bayesian approaches. }}.
\newblock Lecture Notes in Statistics, Springer Verlag, New York.

\bibitem[\protect\citeauthoryear{Ruppert, Wand, and Carrol}{Ruppert
  et~al.}{2003}]{ruppert}
Ruppert, D., Wand, M., and Carrol, R. ({2003}).
\newblock {\em {Semiparametric regression}}.
\newblock Cambridge University Press: Cambridge, UK.

\bibitem[\protect\citeauthoryear{Sayer, Bogard, Buonarosa, LeBlanc, Funkhouser,
  Bao, Blankespoor, and Winkler}{Sayer et~al.}{2011}]{sayer}
Sayer, J., Bogard, S., Buonarosa, M., LeBlanc, D., Funkhouser, D., Bao, S.,
  Blankespoor, A., and Winkler, C. (2011).
\newblock {Integrated Vehicle-Based Safety Systems Light-Vehicle Field
  Operational Test Key Findings Report DOT HS 811 416, Retrieved August 26,
  2015, from \verb?http://www.nhtsa.gov/DOT/NHTSA/NVS/Crash%20Avoidance/Tech?
  \verb?nical%20Publications/2011/811416.pdf?}

\bibitem[\protect\citeauthoryear{Smith, Bailey, and Munford}{Smith
  et~al.}{1993}]{smith}
Smith, D., Bailey, T.~C., and Munford, A. ({1993}).
\newblock {Robust classification of artificial neural networks}.
\newblock {\em {Statistics and Computing}} {\bf {3},} {71--81}.

\bibitem[\protect\citeauthoryear{Tanner and Wong}{Tanner and
  Wong}{1987}]{tanner}
Tanner, M. and Wong, W. ({1987}).
\newblock {The Calculation of Posterior Distributions by Data Augmentation}.
\newblock {\em {Journal of the American Statistical Association}} {\bf {82},}
  {528--540}.

\bibitem[\protect\citeauthoryear{van~der Laan and Polley}{van~der Laan and
  Polley}{2010}]{vanderLaan}
van~der Laan, M. and Polley, E.~C. ({2010}).
\newblock {Super Learner in Prediction}.
\newblock {\em {U.C. Berkeley Division of Biostatistics Working Paper Series}}
  {\bf {Working Paper 266},}
  {\verb?http://biostats.bepress.com/ucbbiostat/paper266?}

\bibitem[\protect\citeauthoryear{Zhang, Shih, and M\"uller}{Zhang
  et~al.}{2007}]{zhang}
Zhang, S., Shih, Y., and M\"uller, P. ({2007}).
\newblock {A Spatially-adjusted Bayesian Additive Regression Tree Model to
  Merge Two Datasets}.
\newblock {\em {Bayesian Analysis}} {\bf {2},} {611--634}.

\end{thebibliography}
\bibliographystyle{biom}

\begin{table}[H]
	\caption{Simulation results for continuous correlated outcomes. Bias and coverage of $\sum_{j=1}^mg(\mathbf{X}_k,T_j,\mathbf{M}_j)+a_k$ $(g(x)+a_k)$, $\sigma$, and $\tau$ for BART, riBART with $P(\tau^2)\propto 1$ (flat), half-Cauchy prior on $\tau^2$, and $\tau^2\sim IG(1,1)$ (proper). \label{biascover_con}}
	\vspace*{0.2cm}
	\hspace*{-1cm}
	\centering
	\tiny
	\begin{tabular}{lcccccccccccccc}	
	\hline
		\multicolumn{14}{l}{\underline{Scenario 1: continuous, $n_k=5$, $K=50$, $\tau=1$, $\sigma=1$}} \\
		 &\multicolumn{4}{c}{$g(x)+a_k$} & & \multicolumn{4}{c}{$\sigma$} & & \multicolumn{4}{c}{$\tau$} \\ \cline{2-5} \cline{7-10} \cline{12-15}
		& \underline{Bias} & \underline{RMSE} &\underline{Coverage (\%)} & \underline{AIL$^*$} & & \underline{Bias} & \underline{RMSE} &\underline{Coverage (\%)} & \underline{AIL} & & \underline{Bias} & \underline{RMSE} &\underline{Coverage (\%)} & \underline{AIL} \\
		BART & $<0.01$ & 0.07 & 95.16 & 3.32 & & 0.13 & 0.18 & 68.50 & 0.40 & & - & - & - & - \\   
		Flat & $<0.01$ & 0.07 & 97.92 & 3.40 & & 0.10 & 0.11 & 91.50 & 0.35 & & -0.11 & 0.19 & 92.50 & 0.65 \\
		Half-Cauchy & $<0.01$ & 0.07 & 97.88 & 3.41 & & 0.11 & 0.12 & 92.00 & 0.36 & & -0.17 & 0.24 & 83.00 & 0.65 \\
		Proper & $<0.01$ & 0.07 & 97.92 & 3.40 & & 0.10 & 0.11 & 92.00 & 0.35 & & -0.13 & 0.18 & 90.50 & 0.57 \\
	\hline
	\multicolumn{14}{l}{\underline{Scenario 2: continuous, $n_k=20$, $K=100$, $\tau=1$, $\sigma=1$}} \\
		 &\multicolumn{4}{c}{$g(x)+a_k$} & & \multicolumn{4}{c}{$\sigma$} & & \multicolumn{4}{c}{$\tau$} \\ \cline{2-5} \cline{7-10} \cline{12-15}
		& \underline{Bias} & \underline{RMSE} &\underline{Coverage (\%)} & \underline{AIL} & & \underline{Bias} & \underline{RMSE} &\underline{Coverage (\%)} & \underline{AIL} & & \underline{Bias} & \underline{RMSE} &\underline{Coverage (\%)} & \underline{AIL} \\
		BART & $<0.01$ & 0.02 & 77.80 & 2.28 & & 0.35 & 0.35 & 0.00 & 0.10 & & - & - & - & - \\   
		Flat & $<0.01$ & 0.02 & 94.25 & 1.81 & & -0.02 & 0.03 & 85.00 & 0.07 & & 0.01 & 0.08 & 92.00 & 0.31 \\
		Half-Cauchy & $<0.01$ & 0.02 & 94.28 & 1.81 & & -0.02 & 0.03 & 84.50 & 0.07 & & $<0.01$ & 0.08 & 92.00 & 0.30 \\
		Proper & $<0.01$ & 0.02 & 94.35 & 1.81 & & -0.02 & 0.02 & 84.00 & 0.07 & & $<0.01$ & 0.08 & 92.00 & 0.30 \\
	\hline
		\multicolumn{14}{l}{\underline{Scenario 3: continuous, $n_k=5$, $K=50$, $\tau=0.5$, $\sigma=1$}} \\
		 &\multicolumn{4}{c}{$g(x)+a_k$} & & \multicolumn{4}{c}{$\sigma$} & & \multicolumn{4}{c}{$\tau$} \\ \cline{2-5} \cline{7-10} \cline{12-15}
		& \underline{Bias} & \underline{RMSE} &\underline{Coverage (\%)} & \underline{AIL} & & \underline{Bias} & \underline{RMSE} &\underline{Coverage (\%)} & \underline{AIL} & & \underline{Bias} & \underline{RMSE} &\underline{Coverage (\%)} & \underline{AIL} \\
		BART & $<0.01$ & 0.07 & 92.54 & 2.66 & & -0.16 & 0.18 & 57.00 & 0.34 & & - & - & - & - \\   
		Flat & $<0.01$ & 0.07 & 97.79 & 3.22 & & 0.07 & 0.09 & 97.00 & 0.33 & & -0.15 & 0.17 & 94.00 & 0.56 \\
		Half-Cauchy & $<0.01$ & 0.07 & 97.77 & 3.21 & & 0.08 & 0.09 & 95.00 & 0.33 & & -0.25 & 0.26 & 72.50 & 0.55 \\
		Proper & $<0.01$ & 0.07 & 97.78 & 3.24 & & 0.06 & 0.08 & 98.00 & 0.32 & & 0.04 & 0.06 & 100.00 & 0.38 \\
	\hline
	\multicolumn{14}{l}{\underline{Scenario 4: continuous, $n_k=20$, $K=100$, $\tau=0.5$, $\sigma=1$}} \\
		 &\multicolumn{4}{c}{$g(x)+a_k$} & & \multicolumn{4}{c}{$\sigma$} & & \multicolumn{4}{c}{$\tau$} \\ \cline{2-5} \cline{7-10} \cline{12-15}
		& \underline{Bias} & \underline{RMSE} &\underline{Coverage (\%)} & \underline{AIL} & & \underline{Bias} & \underline{RMSE} &\underline{Coverage (\%)} & \underline{AIL} & & \underline{Bias} & \underline{RMSE} &\underline{Coverage (\%)} & \underline{AIL} \\
		BART & $<0.01$ & 0.02 & 89.40 & 1.89 & & 0.06 & 0.07 & 12.00 & 0.08 & & - & - & - & - \\   
		Flat & $<0.01$ & 0.02 & 94.27 & 1.79 & & -0.02 & 0.03 & 85.50 & 0.08 & & 0.01 & 0.05 & 93.50 & 0.18 \\
		Half-Cauchy & $<0.01$ & 0.02 & 94.20 & 1.78 & & -0.02 & 0.02 & 85.00 & 0.07 & & $<0.01$ & 0.05 & 92.50 & 0.18 \\
		Proper & $<0.01$ & 0.02 & 94.22 & 1.79 & & -0.02 & 0.03 & 83.50 & 0.07 & & 0.02 & 0.05 & 94.50 & 0.17 \\
	\hline
	\end{tabular}
	
	*AIL = Average interval length.
\end{table}
		
\begin{table}[H]
	\caption{Simulation results for binary correlated outcomes. Bias and coverage of $\sum_{j=1}^mg(\mathbf{X}_k,T_j,\mathbf{M}_j)+a_k$ $(g(x)+a_k)$ and $\tau$ for BART, riBART with $P(\tau^2)\propto 1$ (flat), half-Cauchy prior on $\tau^2$, and $\tau^2\sim IG(1,1)$ (proper). \label{biascover_bin}}
	\vspace*{0.2cm}
	\centering
	\scriptsize
	\begin{tabular}{lcccccccccc}	
	\hline
		\multicolumn{10}{l}{\underline{Scenario 5: binary, $n_k=5$, $K=50$, $\tau=1$}} \\
		 &\multicolumn{4}{c}{$g(x)+a_k$} & & \multicolumn{4}{c}{$\tau$}\\ \cline{2-5} \cline{7-10}
		& \underline{Bias} & \underline{RMSE} &\underline{Coverage (\%)} & \underline{AIL$^*$} & & \underline{Bias} & \underline{RMSE} &\underline{Coverage (\%)} & \underline{AIL} \\
		BART & 0.02 & 0.08 & 66.83 & 1.87 & & - & - & - & - \\  
		Flat & 0.01 & 0.09 & 94.49 & 2.64 & & 0.04 & 0.21 & 94.00 & 0.85 \\
		Half-Cauchy & 0.01 & 0.09 & 94.19 & 2.60 & & $<0.01$ & 0.02 & 95.00 & 0.83 \\
		Proper & 0.01 & 0.09 & 94.17 & 2.57 & & -0.03 & 0.16 & 97.00 & 0.72 \\	
	\hline
	\multicolumn{10}{l}{\underline{Scenario 6: binary, $n_k=20$, $K=100$, $\tau=1$}} \\
		 &\multicolumn{4}{c}{$g(x)+a_k$} & & \multicolumn{4}{c}{$\tau$} \\ \cline{2-5} \cline{7-10}
		& \underline{Bias} & \underline{RMSE} &\underline{Coverage (\%)} & \underline{AIL} & & \underline{Bias} & \underline{RMSE} &\underline{Coverage (\%)} & \underline{AIL} \\
		BART & 0.01 & 0.04 & 45.39 & 1.2 & & - & - & - & - \\   
		Flat & $<0.01$ & 0.04 & 94.87 & 1.58 & & 0.01 & 0.09 & 94.50 & 0.36 \\
		Half-Cauchy & $<0.01$ & 0.04 & 94.83 & 1.58 & & 0.01 & 0.09 & 95.00 & 0.36 \\
		Proper & $<0.01$ & 0.04 & 94.81 & 1.58 & & $<0.01$ & 0.09 & 93.50 & 0.35 \\
	\hline
		\multicolumn{10}{l}{\underline{Scenario 7: binary, $n_k=5$, $K=50$, $\tau=0.5$}} \\
		 &\multicolumn{4}{c}{$g(x)+a_k$} & & \multicolumn{4}{c}{$\tau$} \\ \cline{2-5} \cline{7-10}
		& \underline{Bias} & \underline{RMSE} &\underline{Coverage (\%)} & \underline{AIL} & & \underline{Bias} & \underline{RMSE} &\underline{Coverage (\%)} & \underline{AIL}  \\
		BART & -0.01 & 0.09 & 89.68 & 1.89 & & - & - & - & - \\   
		Flat & -0.01 & 0.09 & 94.78 & 2.06 & & 0.04 & 0.15 & 97.50 & 0.65 \\
		Half-Cauchy & -0.01 & 0.09 & 93.67 & 1.97 & & -0.03 & 0.16 & 96.50 & 0.68 \\
		Proper & -0.01 & 0.09 & 96.03 & 2.17 & & 0.13 & 0.15 & 92.00 & 0.47 \\
	\hline
	\multicolumn{10}{l}{\underline{Scenario 8: binary, $n_k=20$, $K=100$, $\tau=0.5$}} \\
		 &\multicolumn{4}{c}{$g(x)+a_k$} & & \multicolumn{4}{c}{$\tau$} \\ \cline{2-5} \cline{7-10}
		& \underline{Bias} & \underline{RMSE} &\underline{Coverage (\%)} & \underline{AIL} & & \underline{Bias} & \underline{RMSE} &\underline{Coverage (\%)} & \underline{AIL} \\
		BART & $<0.01$ & 0.03 & 74.76 & 1.22 & & - & - & - & - \\   
		Flat & -0.01 & 0.03 & 94.83 & 1.35 & & 0.01 & 0.05 & 95.50 & 0.21 \\
		Half-Cauchy & -0.01 & 0.03 & 94.72 & 1.34 & & $<0.01$ & 0.05 & 94.50 & 0.21 \\
		Proper & -0.01 & 0.03 & 95.01 & 1.36 & & 0.03 & 0.05 & 95.00 & 0.2 \\ 
	\hline
	\end{tabular}
	
	*AIL = Average interval length.
\end{table}

\end{document}